\documentclass[pdf-a,balance,colorlinks,upint,subscriptcorrection,varvw,mathalfa=cal=boondoxo, spanish,french,vietnamese,russian,greek]{asmeconf}
\hypersetup{%
	pdfauthor={Sreetam Bhaduri},
	pdftitle={ASME IMECE 2025 SB V1},
	pdfkeywords={ASME,IMECE,Sreetam},
	pdfsubject = {Asymptotic Behavior of a Buoyant Jet Regime inside a Carbon-dioxide Ejector},
}
\usepackage{multirow}
\begin{document}

\ConfName{Proceedings of the ASME 2025\linebreak International Mechanical Engineering Congress \& Exposition (IMECE)}

\ConfAcronym{IMECE2025}

\ConfDate{November 16–20, 2025}

\ConfCity{Memphis,TN}

\PaperNo{IMECE2025-165694}

\title{Asymptotic Behavior of a Buoyant Jet Regime inside a Carbon-dioxide Ejector}

\SetAuthors{
	Sreetam\ Bhaduri\affil{}\CorrespondingAuthor{}, 
	Ivan C. Christov\affil{}\CorrespondingAuthor{}, 
	Eckhard A. Groll\affil{}\CorrespondingAuthor{},  
	Davide Ziviani\affil{}\CorrespondingAuthor{bhaduri@purdue.edu, christov@purdue.edu, groll@purdue.edu, dziviani@purdue.edu}
	}

\SetAffiliation{}{School of Mechanical Engineering, Purdue
University, West Lafayette, Indiana, United States}

\maketitle

\versionfootnote{Accepted in the 31st Symposium on Fundamental Issues and Perspectives in Fluid Mechanics, 2025 International Mechanical Engineering Congress \& Exposition (IMECE). The \color{red}link \color{black}will be provided after publication.}



\begin{abstract}
Ejectors are used in various engineering systems, including steam and vapor compression cycles. Optimizing the performance of ejectors requires understanding and analysis of multiphase and turbulent flow structures associated with their internal flow fields. This approach yields higher fidelity but at a high computational cost. Lower-fidelity one-dimensional (1D) models offer lower computational costs; however, 1D models are often empirical and provide limited understanding of the internal flow fields, overlooking possibilities of optimization. Ejector flows can be categorized into four regimes: Regime 1 (R1), which is compressibility dominated; Regime 2 (R2), which is interface instability driven; Regime 3 (R3), which is buoyancy dominated; and Regime 4 (R4), which is a wall-bounded turbulent jet expansion. Among these, the buoyancy-dominated regime is the most complex and least understood. This work discusses an approach to develop a reduced-order model utilizing a self-similarity framework to capture the internal flow field of the jet within the buoyancy-dominated regime under quasi-steady, compressible, and isothermal flow conditions, where density variations arise only from mixing. The density variation is captured through the Favre-averaging approach. The model captures the expansion of a central jet influenced by momentum diffusivity and a constant streamwise pressure gradient. Interaction of the central jet with the cylindrical wall induces a counterflow annular wall jet due to the combined effects of negative radial density gradients and shear stress imposed by the wall. Initially, the discussion focuses on flow topology inside the ejector, followed by the self-similarity methodology and implementation of asymptotic analysis. Finally, the resemblance of the self-similar nature of the flow field and the existence of inner and outer regions in the flow field are discussed. The boundary conditions are derived from a validated high-fidelity three-dimensional (3D) numerical simulation of a subcritical liquid-gas carbon dioxide ($\text{CO}_{2}$) ejector used in multi-stage refrigeration. The 3D simulations modeled turbulence via the Reynolds Averaging approach, incorporating a Reynolds Stress Model - Shear Stress Gradient (RSM-SSG) as the closure. The self-proposed similarity methodology developed from the knowledge gained from the 3D simulations enables the development of a low-order model that accurately predicts jet interface characteristics, providing a fast and efficient tool for optimizing ejector designs in cycle-level applications.
\end{abstract}

\section{Introduction} \label{sec:introduction}
The investigation of near-critical state fluid jets is an important problem for various engineering applications such as propulsion and thermal systems. In this context, ejectors are utilized to convert flow work into kinetic energy and ultimately into a pressure lift in various systems, including gas turbines \cite{zhang2013}, liquid propulsion systems \cite{chao2020mixing}, and refrigeration systems \cite{li_transcritical_2005}. The ejector operating principle relies on a high-speed jet in single- or multiphase conditions \cite{croquer_large_2022,bhaduri2024flow}. The efficiency of ejector devices depends on the physics of the jet, especially under multiphase operations \cite{ringstad_detailed_2020,wilhelmsen2022one}. Ejector components are used in various engineering applications as expansion recovery devices. Specifically, ejectors are a flow device that converts kinetic energy into pressure recovery \cite{bodys_non-equilibrium_2020}. Different types of ejectors exist based on the applications, viz.\ ejector used in gas turbine cooling expands high-pressure gas with mixing gas-phase fluid, which increases volumetric efficiency of the combustor \cite{zhang2013}. Meanwhile, the ejector used in refrigeration systems expands the liquid phase of fluid by mixing a gas-phase fluid, developing a two-phase fluid, and decreasing compressor work input \cite{singh_studies_2018}.

\begin{figure}[h]
    \centering
    \includegraphics[width=1.0\linewidth]{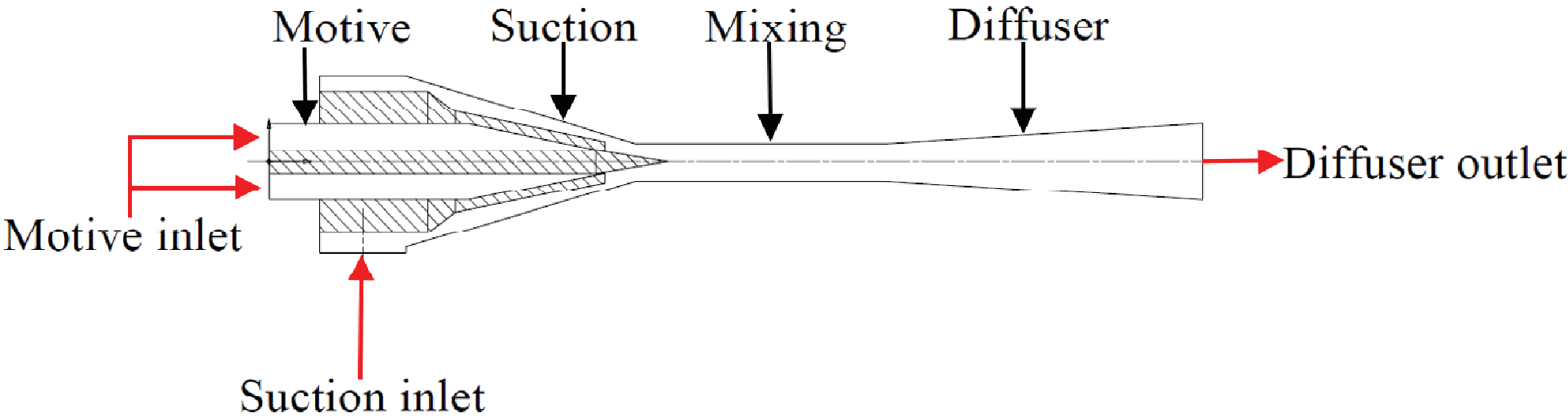}
    \caption{Schematics of an ejector}
    \label{fig:Domain.png}
\end{figure}

Figure \ref{fig:Domain.png} represents a schematic diagram of a variable nozzle inlet ejector. The inlet of an ejector, often known as the `motive', contains liquid at high pressure, and the `suction' includes a vapor phase of the same or different fluid. High-pressure liquid that flows out of the motive throat induces a negative pressure gradient in the suction throat by increasing kinetic energy, which develops a suction effect on the vapor flowing through the suction inlet. These distinct vapor and liquid flows mix downstream through the mixing zone and expand in the following diffuser zone, increasing the pressure \cite{li_transcritical_2005}.

Ejectors are not only used in refrigeration systems, but they are widely applied in oil and natural gas systems for waste gas recovery process \cite{dutton_optimized_1983}, and in gas turbines for their cooling performance increment by improving compressor entrainment efficiency \cite{freedman_novel_1994}. Palacz et al.~\cite{palacz_cfd-based_2016} suggested that the performance of ejector depends on its geometry and operating conditions, and its efficiency could be defined mathematically as follows \cite{elbel_ejector_2008}:
\begin{equation}
    \eta_{ej}=\left(\frac{\dot{m}_{s}}{\dot{m}_{m}}\right)\left(\frac{h_{P_{d},s_{s}}-h_{s}}{h_{m}-h_{P_{d},s_{m}}}\right)
\end{equation}
where $\dot{m}_{s}$ and $\dot{m}_{m}$ are the mass flow rates through the suction and motive regions, respectively, $h_{P_{d},s_{s}}$ is the isentropic (suction) enthalpy at the diffuser, $h_{P_{d},s_{m}}$ is the isentropic (motive) at the diffuser, $h_{s}$ and $h_{m}$ are the actual enthalpy at the suction and motive, respectively. Increasing efficiency demands optimization of the geometry and operating conditions, in particular with respect to the flow characteristics inside the ejector. 

Thermodynamically, ejectors can be understood as a modified expansion device, with a primary assumption of instantaneous mixing by instantaneous motive jet breakup \cite{li_transcritical_2005}. However, local exergy analysis of a liquid-gas $\text{CO}_{2}$ ejector by Wilhelmsen et al.~\cite{wilhelmsen2022one} in 2022 showed the complexity of the physics associated with the mixing zone, where the exergy destruction is 2.3 times higher than that of the diffuser zone. High exergy destruction is related to the entrainment and mixing along the shear layer between the motive and suction fluids. Therefore, a high-fidelity physics-based model capturing the shear layer instability is essential for optimizing the ejector efficiency and increasing entrainment. On the other hand, our previous several analyses \cite{bhaduri2024flow,10.1063/5.0278015} pointed out an increased computational cost for carrying out high-fidelity three-dimensional simulations of the validated numerical model, indicating that optimization via 3D simulations is a slow and tedious process. This observation demonstrates the need for a reduced-order model of an ejector. However, such a model should still incorporate the detailed understanding of the physics of the flow field inside the mixing zone gained from 3D simulations.

Previously \cite{10.1063/5.0278015}, we provided a detailed description of the physics of internal flow-field in the mixing zone of a liquid-gas $\text{CO}_{2}$ ejector, and categorized different regimes of the jet based dominant physics, like, compressibility (R1), interface instability (R2), negatively buoyant jet (R3), and wall-bounded expanding turbulent flow (R4). Model reduction of R1 depends on the isothermal expansivity at an average of motive ($P_{m}$) and suction ($P_{s}$) inlet pressures, and at an average density of motive ($\rho_{m}$) and suction ($\rho_{s}$) inlet densities. The model reduction in R2 takes into account the Kelvin-Helmholtz Instability (KHI) mechanism; its length scales depend on the integral multiples of critical wavelengths. Regime 2 transfers kinetic energy from the motive jet to the surrounding suction flow stream, quiescent without the motive jet. The most complex regime of the jet is R3 before transforming into a wall-bounded expanding turbulent flow in R4. Regime 3 entrains surrounding suction fluid into the motive fluid and mixes, maximizing exergy destruction by mixing. The physics of the internal flow in R3 closely resembles a negatively buoyant jet, where a constant positive pressure gradient replaces the streamwise effect of gravity. The length-scales of this regime are the momentum length, $l_{m}$, of an equivalent negatively buoyant jet, and the expansion angle $\theta_{\text{R3}}$, is a function of the diameter of the mixing zone, $d_{\text{mix}}$. However, model reduction should also capture the physics of mixing along the shear layer, so that entrainment can be incorporated, completing the reduced-order model of R3.

This work provides a thorough discussion on the development of a reduced-order model of R3 in an ejector. Due to its local, and potentially incomplete, self-similar nature, the model reduction is approached by developing governing equations and a self-similar transformation of the flow field, which in the future will be solved using an asymptotic expansion approach.

\section{Numerical Modeling} \label{sec:model}

\subsection{Problem description} \label{sec:problem_description}

\begin{figure}[htbp]
    \centering
    \includegraphics[width=\linewidth]{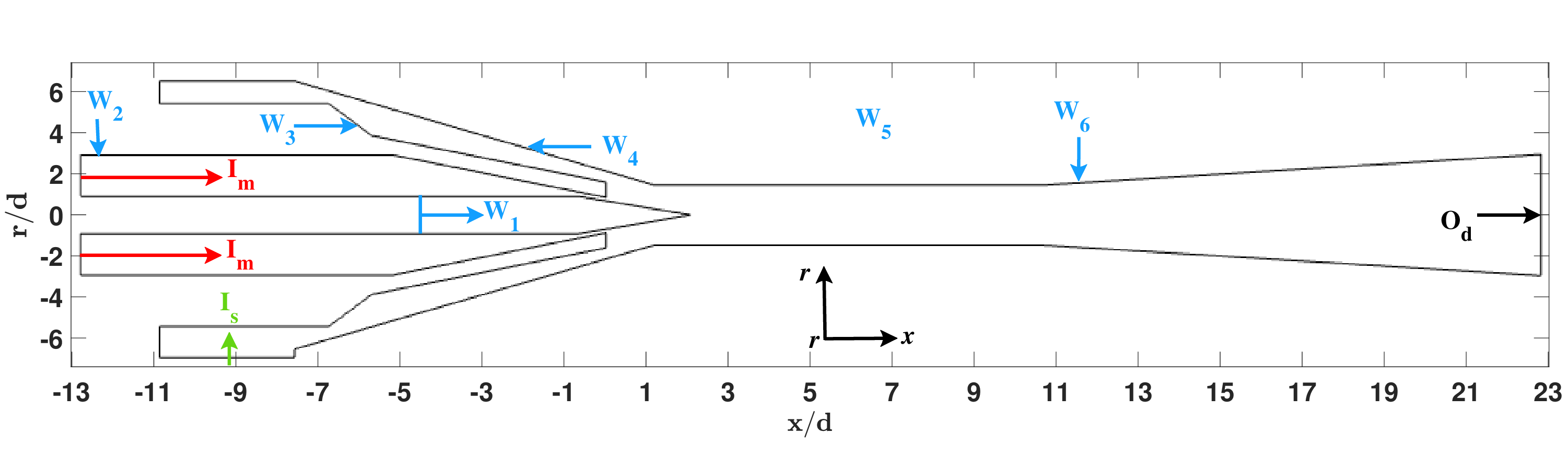}
    \caption{Schematic of the computational (flow) domain extracted from the actual geometry of the ejector. $\text{I}_{\text{m}}$, $\text{I}_{\text{s}}$, $\text{O}_{\text{d}}$, and $\text{W}_{i=[1,6]}$ denote the set of implemented boundary conditions in the domain within $(x, r, \theta)$ space. The velocity vector's components in this coordinate system are $u_{x}$, $u_{r}$, $u_{\theta}$. Here, $\text{I}_{\text{m}} = [P_{m},T_{m},Y_{\text{CO}_{2}(\text{g})}=0]$, $\text{I}_{\text{s}} = [\dot{m}_{s},T_{s},Y_{\text{CO}_{2}(\text{g})}=1]$, $\text{O}_{\text{d}} = P_{d}$, and $\text{W}_{i=[1,6]} = u_{x,\text{wall}}$}
    \label{fig:Problem_def.png}
\end{figure}

The computational flow domain of the ejector is illustrated in Fig.~\ref{fig:Problem_def.png}. The cylindrical coordinate system defines $x$ as the axial direction, $r$ as the radial direction, and $\theta$ as the azimuthal direction. This longitudinal slice through the ejector centerline shows the motive inlet (red, $x/d=-13$), the suction inlet (green, $-11\leq x/d\leq -8$), diffuser outlet (black, $x/d=23$), and bounding walls (cyan). The corresponding velocity vector components are defined by $u_{x}$, $u_{r}$, $u_{\theta}$. The origin is placed at the center of the motive throat. A converging needle controls the inlet opening. The inlet flow is annular. Table~\ref{tab:Boundary conditions} reports the 3D simulations' boundary conditions.

\begin{table*}
    \caption{\label{tab:Boundary conditions}Imposed boundary conditions in the 3D numerical simulations}
    \renewcommand{\arraystretch}{1.3}
    \centering
    \begin{tabular}{p{3cm} p{6cm} p{6.75cm}}
        \toprule
        Region & Inputs from experiment & Imposed boundary conditions \\ 
        \midrule
        Motive inlet &
        \begin{tabular}[t]{@{}l@{}}
            $P_{m} = 63.5$ bar \\
            $\dot{m}_{m} = 57.4$ g/s \\
            $T_{m} = 296$ K \\
            $Y_{\text{CO}_{2}(g)} = 0$ \\
            $\mathcal{H} = 0.38$ mm
        \end{tabular} &
        \begin{tabular}[t]{@{}l@{}}
            $P_{m} = 63.5$ bar \\ 
            $\nabla{\vec{u}} = \vec{0}$ \\ 
            $T_{m} = 296$ K \\ 
            $Y_{\text{CO}_{2}(g)} = 0$ \\
            $\mathcal{H} = 0.31$ mm
        \end{tabular} \\
        \midrule
        Suction inlet &
        \begin{tabular}[t]{@{}l@{}} 
            $P_{s} = 25.8$ bar \\
            $\dot{m}_{s} = 9.85$ g/s \\
            $T_{s} = 279$ K \\ 
            $Y_{\text{CO}_{2}(g)} = 1$
        \end{tabular} &
        \begin{tabular}[t]{@{}l@{}}
            $\nabla p = \vec{0}$ \\ 
            $\dot{m}_{s} = 9.85$ g/s \\ 
            $T_{s} = 279$ K \\ 
            $Y_{\text{CO}_{2}(g)} = 1$
        \end{tabular} \\
        \midrule
        Diffuser outlet &
        \begin{tabular}[t]{@{}l@{}} 
            $P_{d} = 30.3$ bar \\ 
            $\dot{m}_{d} = 67.2$ g/s \\
            $Y_{\text{CO}_{2}(g)} = x_{d} = 0.43$
        \end{tabular} &
        \begin{tabular}[t]{@{}l@{}}
            $P_{d} = 30.3$ bar \\ 
            $\nabla\vec{u} = \vec{0}$ \\ 
            No backflow
        \end{tabular} \\
        \midrule
        Walls &
        Not measured &
        \begin{tabular}[t]{@{}l@{}}
            $u_{x,\text{wall}} = 0$ \\ 
            Wall motion: Stationary and fixed \\ 
            $y^{+}$: 0.8 (on the needle wall), 1 (all other walls) \\ 
            Absolute roughness: 0.01 mm \\ 
            Roughness constant: 0.5 \\ 
            Near-wall treatment: Standard law wall model
        \end{tabular} \\
        \bottomrule
    \end{tabular}
\end{table*}

\subsection{Governing equations} \label{sec:governing_equations}
The three-dimensional compressible Navier-Stokes equations are coupled to multiphase transport equations to simulate the flow inside the fluid domain. The continuity, momentum, energy, species, and void fraction transport equations are applied to any thermodynamic state of $\text{CO}_{2}$ \cite{pope_turbulent_2000,bilicki1990physical}. Turbulence is modeled using the Reynolds Averaged Navier-Stokes (RANS) approach, with a turbulence closure based on the Reynolds Stress Model-Shear Stress Gradient (RSM-SSG) \cite{bhaduri2024flow}. The thermophysical and transport properties of $\text{CO}_{2}$ as a function of pressure and temperature are taken from the National Institute of Standards and Technology (NIST) database \cite{lemmon2010nist}. The Peng-Robinson equation of state (EoS) is employed to obtain real-fluid properties. No body force is present in this problem due to its relatively small length scales. Homogeneous Relation Model (HRM) is applied to capture the physics of phase change, assuming independent condensation and evaporation time scales, $\theta_{\text{c}}$ and $\theta_{\text{e}}$, respectively, where $\theta_{\text{c}}\neq \theta_{\text{e}}$ \cite{bilicki1990physical,saha_investigation_2017,richards_converge_2023}. High-Resolution Interface Capture (HRIC) was used to better resolve sharp interfaces in this multiphase flow \cite{waclawczyk_comparison_nodate}.

\subsection{Numerics}
We performed 3D transient simulations with a variable time step algorithm allowing minimum and maximum time step sizes of 10 ns and 1 ms, respectively, limiting the change in time step size to 25\% of the previous time step size as the solution advances \cite{richards_converge_2023}. The pressure-velocity coupling is achieved using the Pressure Implicit Splitting Order (PISO) algorithm \cite{tukovic_consistent_2018}, the pressure Poisson equation is discretized with a HYPRE Bi-Conjugate Stabilized (BiCGSTAB) scheme \cite{jan_quasi-implicit_2007}, and the discretized momentum and species transport equations are solved with the Successive Over Relaxation (SOR) scheme with an over-relaxation factor of 1.0. The time step size is calculated based on convective, diffusive, and compressibility stability limits according to a Courant-Friedrichs-Lewy (CFL) number. Specifically, for convection $\text{CFL}_{c} = \vert\vec{u}\vert\Delta t/\Delta$ and we require  $\text{CFL}_{c} \leq 0.5$. Meanwhile, the CFL-like number for diffusion is $\text{CFL}_{d} = (\nu + \nu_{\text{T}}) \Delta t/\Delta^{2} \leq 0.25$, and the CFL-like number for compressibility is $\text{CFL}_{a} = a\Delta t/\Delta \leq 5$. From the three constraints, the most restrictive constraint is enforced in the simulations \cite{richards_converge_2023}. Here, $\nu$ is the local momentum diffusivity, $\nu_{\text{T}}$ is the local turbulent momentum diffusivity, $\Delta$ is the local mesh cell edge length, $a$ is the local speed of sound, and $\Delta t$ is the global time step size. The variable time step algorithm is based on the grid size controlled by Adaptive Mesh Refinement (AMR), the number of iterations at the previous time step, and the CFL numbers. 

The flow domain is discretized with a Cartesian grid generated using a cut cell method implemented in the CONVERGE CFD software \cite{richards_converge_2023}. The grid is a Cartesian grid with uniform edge lengths in the three coordinate directions. The baseline grid, $\Delta_{\text{base}}$,  for this study has $\Delta_{\text{base}} = 4$ mm grid resolution based on our grid convergence study \cite{10.1063/5.0278015}. To avoid cell defeaturing and loss of orthogonality at constricted geometry, the base grid is refined at these locations to a refinement level of $n=5$, minimizing local grid size as $\Delta = \Delta_{\text{base}}/2^{n}$ to $0.125 \ \text{mm}$ \cite{bhaduri2024flow}. AMR is also implemented, which depends on the gradients as listed in Table~\ref{tab:Grid}, adapting the computational mesh to the evolving features of the flow and species fields.

\begin{figure}[htbp]
    \centering
    \includegraphics[width=\linewidth]{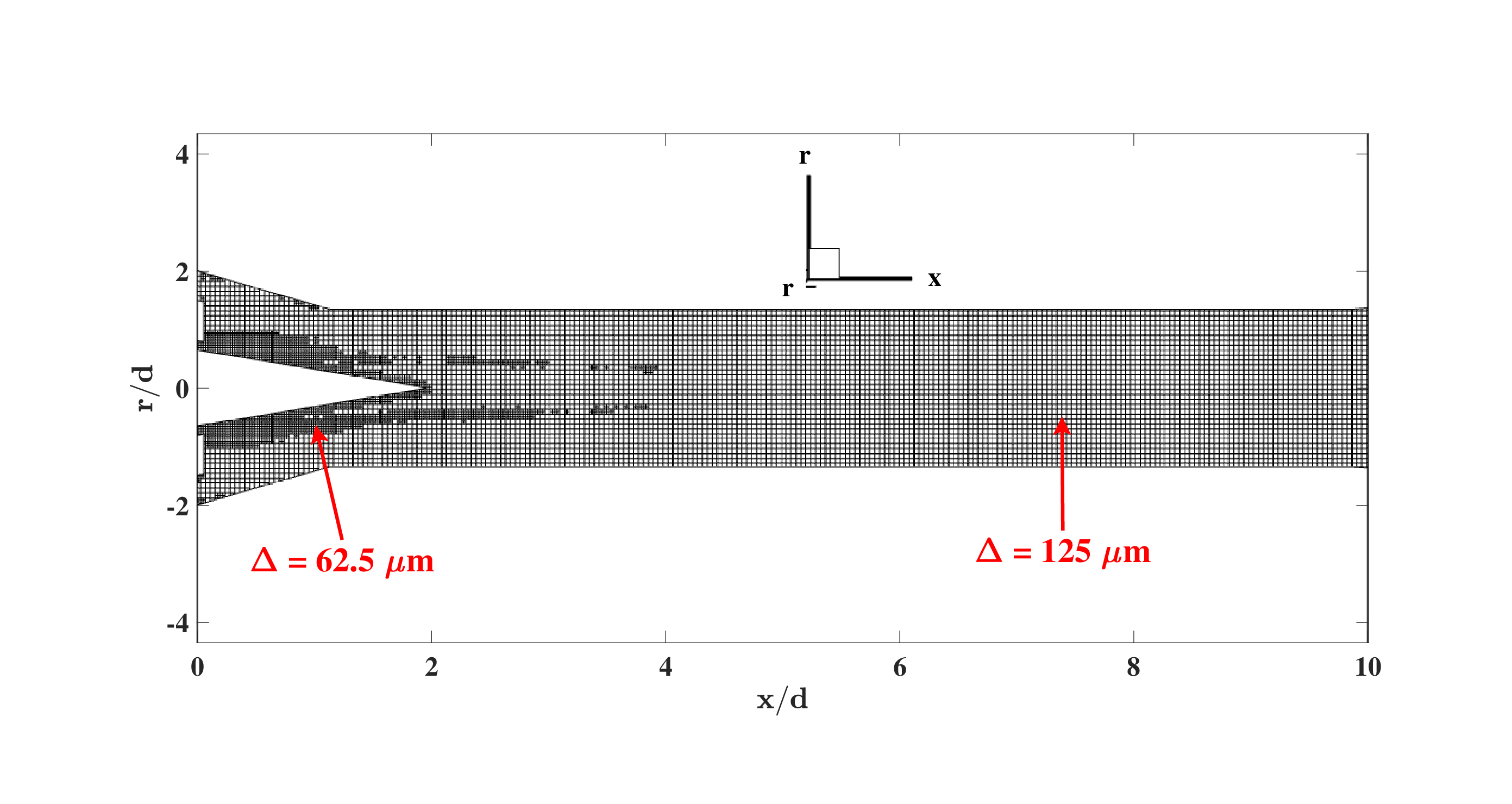}
    \caption{\label{fig:Grid.png}Illustration of the spatial distribution of the grid in mixing zone at $t^{*}=t/t_{\text{end}} = 1$}
\end{figure}

\begin{table}[htbp]
    \caption{\label{tab:Grid}Details of the grid implemented with $\Delta_{\text{base}} = 4.0 $ mm and adaptive mesh refinement}
    \centering
    \renewcommand{\arraystretch}{1.3} 
    \begin{tabular}{>{\centering\arraybackslash}p{0.25\linewidth} 
                    >{\centering\arraybackslash}p{0.03\linewidth} 
                    >{\centering\arraybackslash}p{0.15\linewidth} 
                    >{\arraybackslash}p{0.35\linewidth}} 
    \hline
    Region & $n$ & $\Delta$ ($\mu$m) & Refinement condition \\ \hline
    Entire domain & 5 & 125 & $\partial\Delta/\partial t = 0$ \\ \hline
    \multirow{3}{*}{AMR} & \multirow{3}{*}{6} & \multirow{3}{*}{62.5}
            & $|\nabla \vec{\bar{u}}| > 0.1 $ 1/s \\
            & & & $|\nabla\bar{T}| > 0.5 $ K/m \\
            & & & $|\nabla\bar{\alpha}| > 10^{-5}$ 1/m \\ \hline
    \end{tabular}
\end{table}

These discretization strategies lead to a more accurate solution, capturing complex flow and topological changes, compared to standard interpolation schemes \cite{jan_quasi-implicit_2007,capuani_discrete_2004}.  Convergence criteria are based on the scaled momentum, energy, and species equations' residuals, which are maintained at $\le 10^{-5}$. Performing a grid convergence test, the uncertainties meet the convergence criteria, proving that the governing equations are solved correctly \cite{10.1063/5.0278015}.

\section{Validation and Verification} \label{sec:validation}
The numerical model from the previous section was validated against measured data from the experimental facility at the Ray W.\ Herrick Laboratories \cite{bhaduri2024flow}. Table~\ref{tab:Boundary conditions} summarizes the boundary conditions in the motive region, where Dirichlet-type pressure, temperature, and Neumann-type velocity boundary conditions are provided, and the simulation solves for the mass flow rate. Similarly, for the suction region, a Dirichlet-type mass flow rate, temperature, and Neumann-type pressure boundary conditions are provided, and the code solves for the pressure.

\begin{figure*}[htbp]
    \parbox[c]{\linewidth}{
    \centering
    \begin{subfigure}[htbp]{0.49\linewidth}
        \centering
        \includegraphics[width=\linewidth]{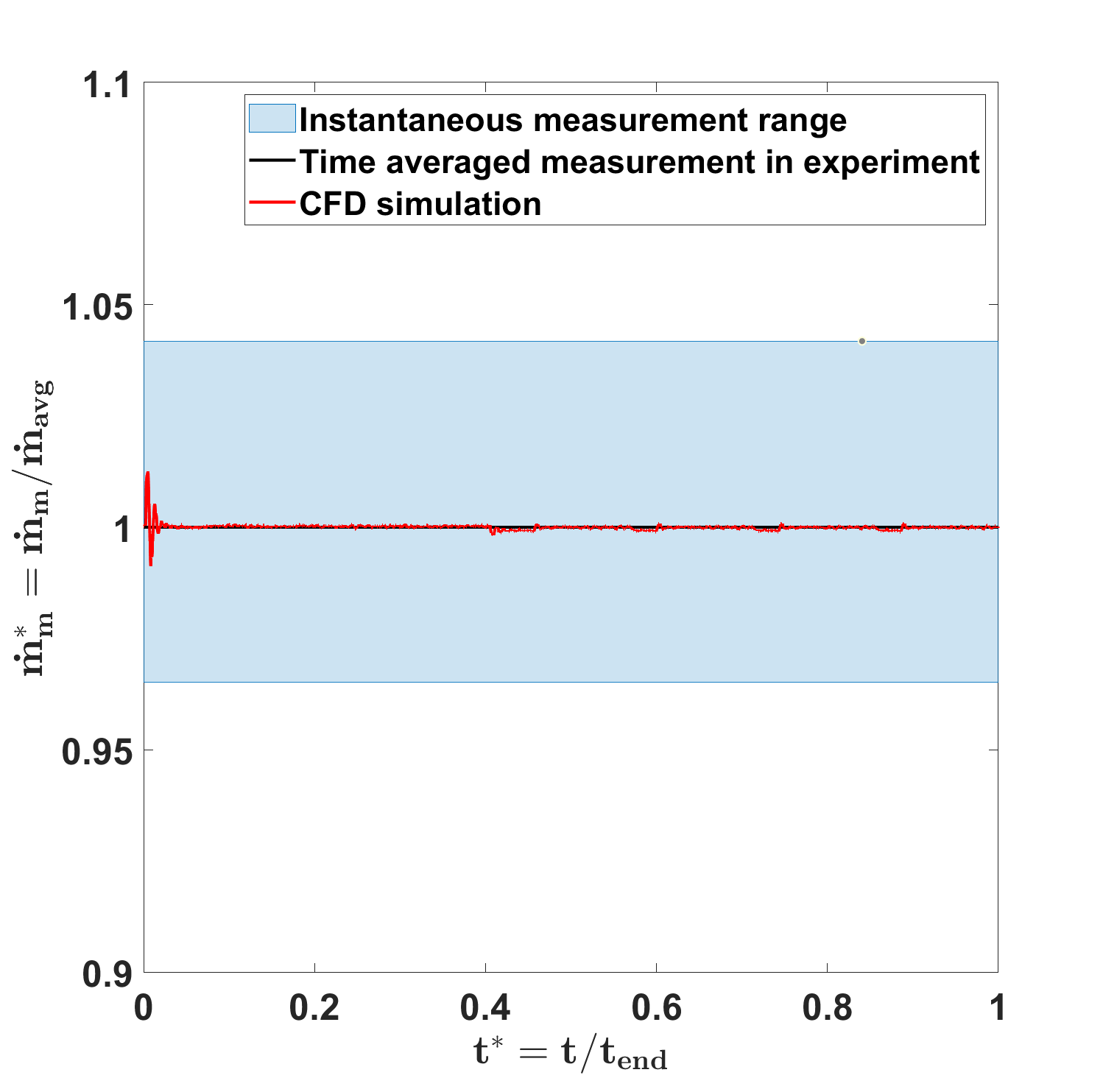}
        \caption{$\dot{m}^{*}=\dot{m}_{\text{m}}/\dot{m}_{\text{avg}}$}
    \end{subfigure}
    \begin{subfigure}[htbp]{0.49\linewidth}
        \centering
        \includegraphics[width=\linewidth]{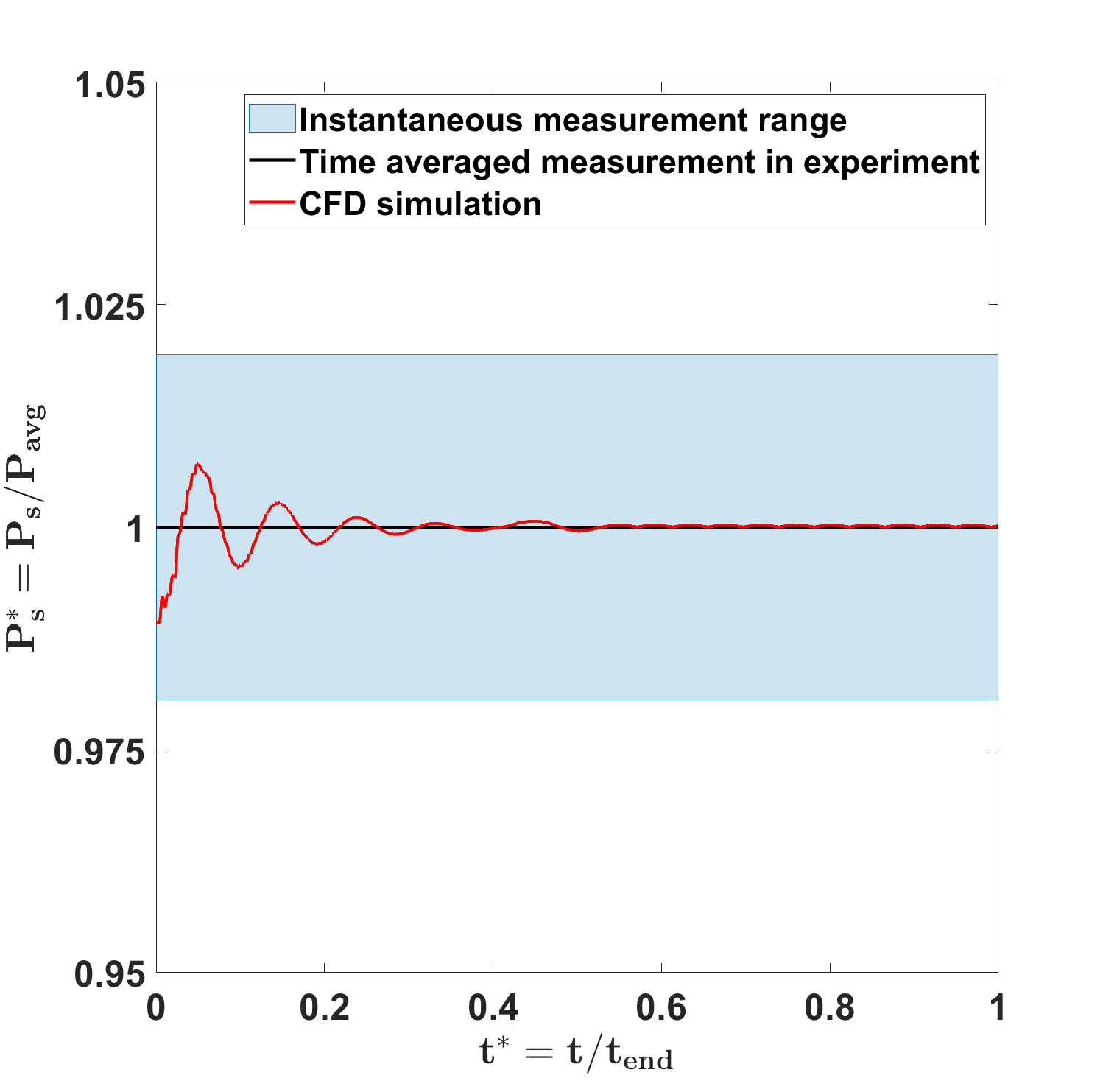}
        \caption{$P^{*}=P_{\text{s}}/P_{\text{avg}}$}
    \end{subfigure}
    \caption{Comparison of simulated normalized mass flow rate and pressure with measurements from experiments}
    \label{fig:Validation.png}}
\end{figure*}

Figure~\ref{fig:Validation.png} shows a comparison of the simulated motive mass flow rate $\dot{m}_{m}$ and suction pressure $P_{s}$ with the corresponding measured quantities. Table~\ref{tab:Time averaged comparison} quantitatively compares the time-averaged simulated quantities with time-averaged measured quantities as a function of Root Mean Square Error (RMSE). The numerical simulation under-predicted the motive mass flow rate by only  $\Delta\dot{m}_{m} = 0.84\%$, and the suction pressure by only $\Delta P_{s} = 0.54\%$, which gives us confidence in the simulations.

\begin{table}[htbp]
    \centering
    \caption{\label{tab:Time averaged comparison}Comparison of time-averaged simulated quantities with time-averaged measured quantities}
    \resizebox{\linewidth}{!}{
    \begin{tabular}{c c c c c c c}
        \hline
        Method & $\dot{m}_{m}$ & $\Delta\dot{m}_{m}$ & $P_{s}$ & $\Delta P_{s}$ & $Y_{\textbf{CO}_{2}(g)}$ & $\Delta Y_{\textbf{CO}_{2}(g)}$ \\
        -- & (g/s) & (\%) & (bar) & (\%) & (\%) & (\%) \\ \hline
        Sim. & 56.88 & $-0.84$ & 25.92 & 0.54 & 42.82 & $-0.23$ \\ \hline
        Exp. & 57.36 & - & 25.78 & -- & 42.92 & -- \\ \hline
    \end{tabular}}
\end{table}

The uncertainty associated with instabilities during the experiment is represented in Fig.\ref{fig:Validation.png} as a blue (shaded) region. The uncertainty originates from the compressor fluctuations, which provide a standard deviation of $0.69\%$ of the measured quantity \cite{bhaduri2024flow}. Clearly, the 3D numerical simulations predicted the temporal evolution of $\dot{m}_{m}$ and $P_{s}$ within the measured uncertainty ranges.

\section{Internal Flow Topology} \label{sec:flow}
\begin{figure*}[htbp]
    \parbox[c]{\linewidth}{
    \centering
    \begin{subfigure}[htbp]{\linewidth}
        \centering
        \includegraphics[width=\linewidth]{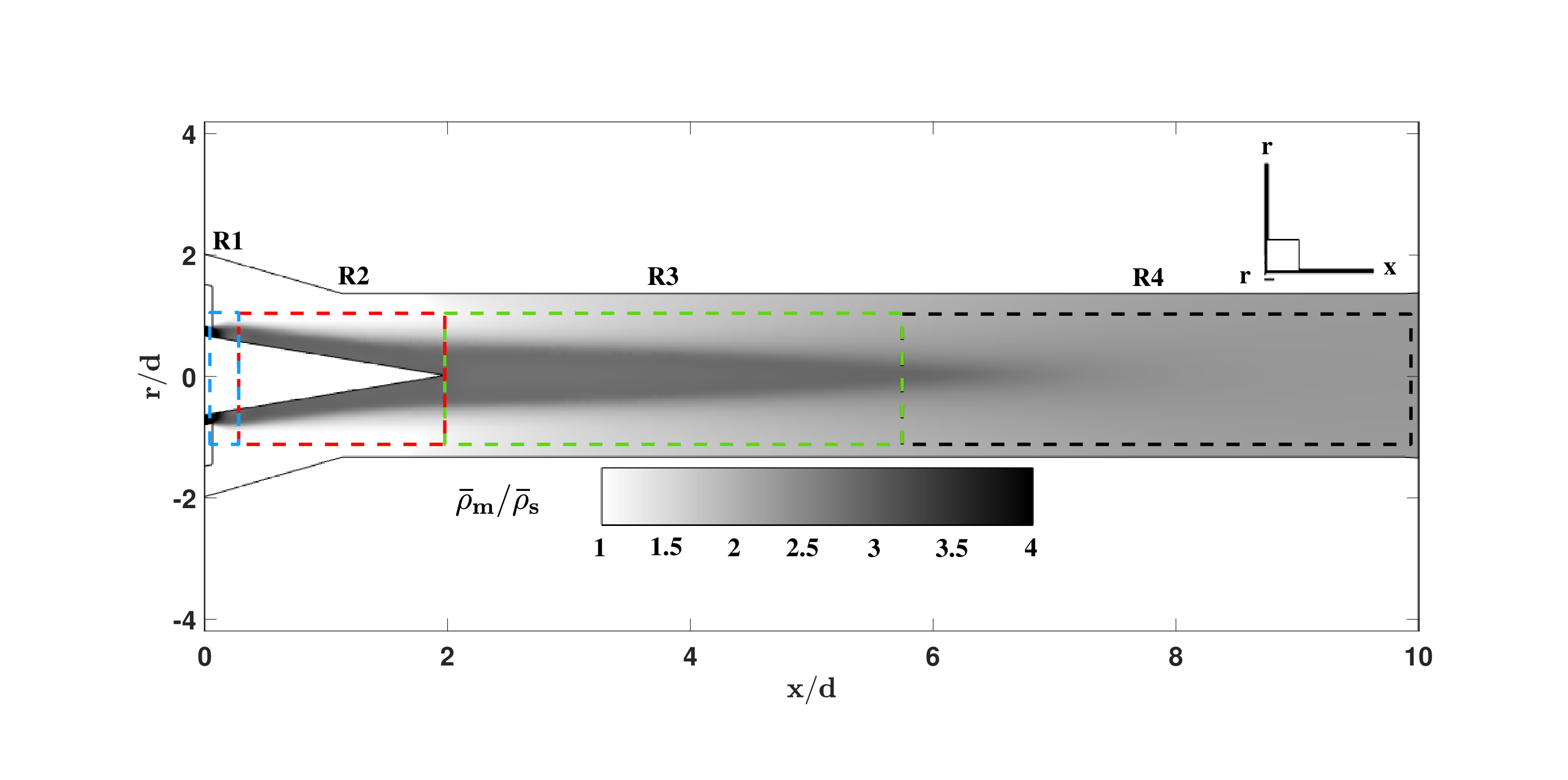}
        \caption{Spatial distribution}
    \end{subfigure}
    \begin{subfigure}[htbp]{0.49\linewidth}
        \centering
        \includegraphics[width=\linewidth]{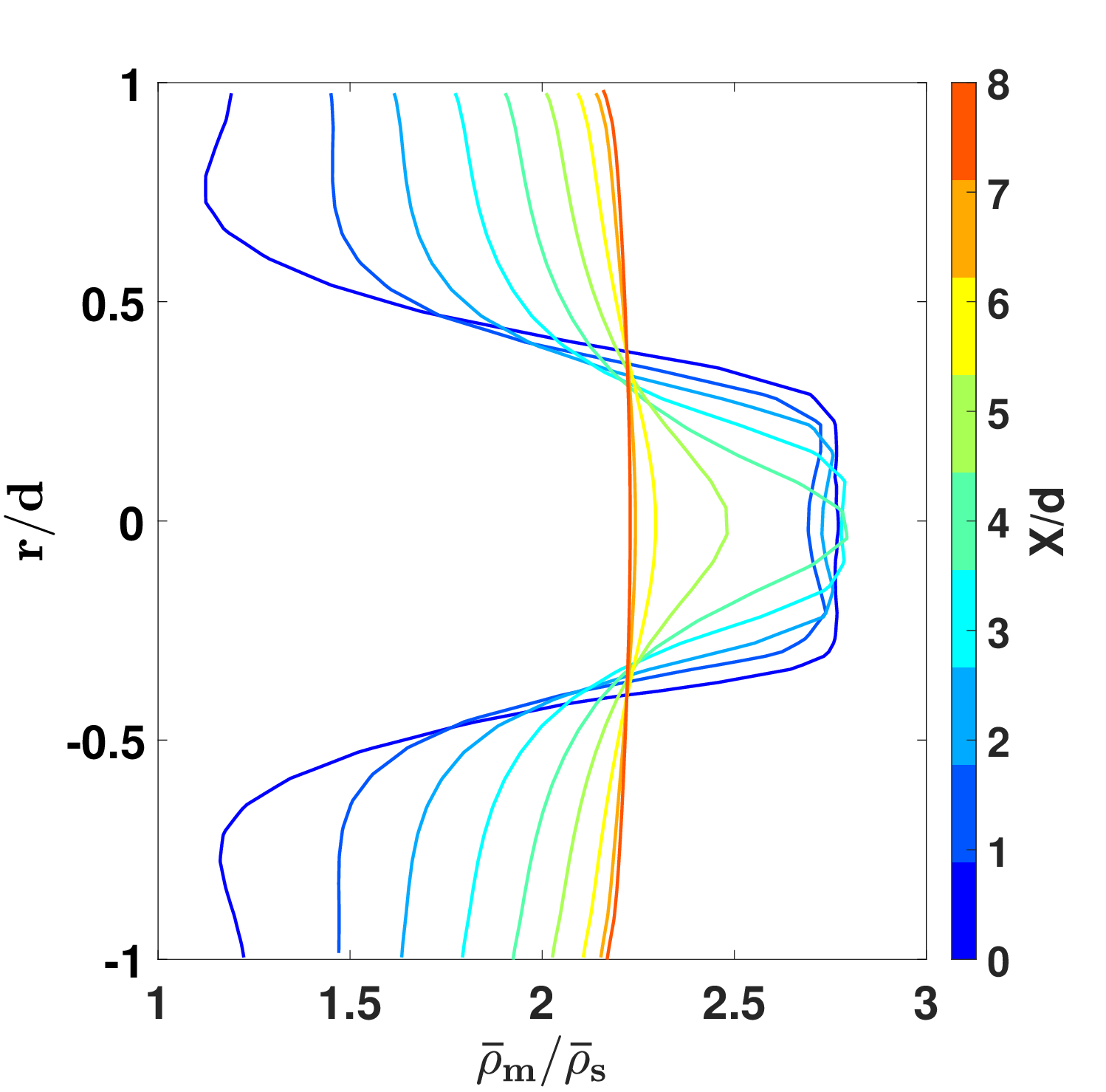}
        \caption{Radial distribution for $0 \leq X/d \leq 8$}
    \end{subfigure}
    \hfill\hfill
    \begin{subfigure}[htbp]{0.49\linewidth}
        \centering
        \includegraphics[width=\linewidth]{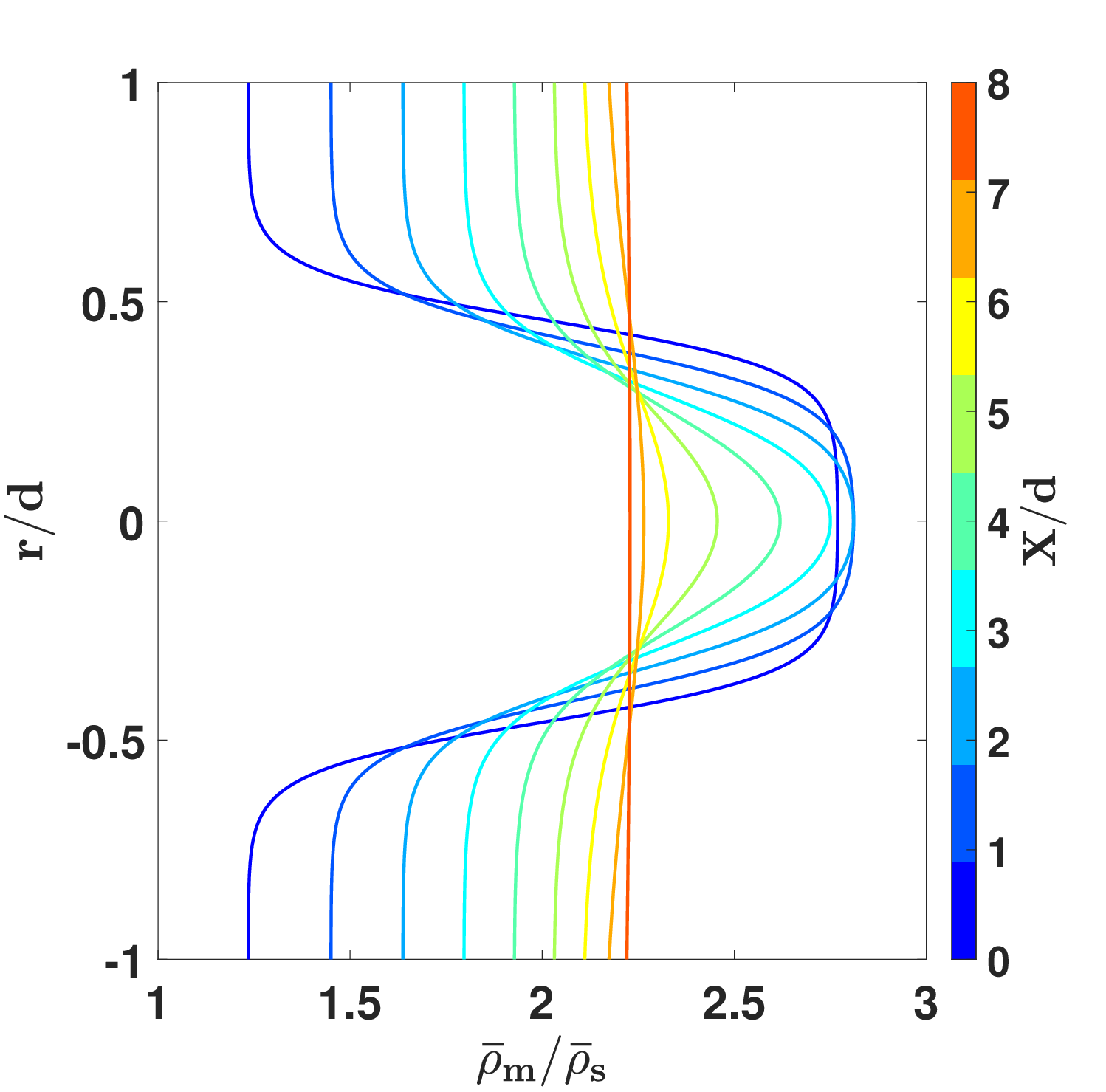}
        \caption{Fitted radial distribution for $0 \leq X/d \leq 8$}
    \end{subfigure}
    \caption{Spatial and radial distributions of the normalized time-averaged mixture density $\bar{\rho}_{m}/\bar{\rho}_{s}$ from (a) numerical simulation and (b) fitted to the analytical function in Eq.~\eqref{equ:sample_fit}}
    \label{fig:contour.png}}
\end{figure*}

Different regimes of the motive jet are primarily responsible for the entrainment of surrounding suction fluid into the motive fluid at the expense of phase change and mixing along the shear layer. These regimes are characterized by different physics, such as compressibility (regime R1), interface instability (regime R2), destabilization of core flow (regime R3), and simple turbulent flow (regime R4) acting concurrently inside the ejector \cite{10.1063/5.0278015}. The spatial and radial distributions of normalized time-averaged mixture density, $\bar{\rho}_{m}/\bar{\rho}_{s}$, of the jet identified with different regimes are illustrated in Fig.~\ref{fig:contour.png}(a). Note that the time-averaged mixture density $\bar{\rho}_{m}$ is normalized using the time-averaged suction inlet density $\bar{\rho}_{s}$.

Figure \ref{fig:contour.png}(a) shows that the liquid through the annular motive throat develops a jet with different regimes from R1 to R4 inside the mixing zone. The time-averaged mixture density, $\bar{\rho}_{m}$, decreases in the axial direction until the termination of R3. Due to mixing across the shear layer of the jet, the ratio $\bar{\rho}_{m}/\bar{\rho}_{s} \rightarrow 2$ at the end of R3. This observation indicates a decreasing radial density gradient along the axial direction, which is confirmed by Fig.~\ref{fig:contour.png}(b). Here, $X = (x - l_{\text{R2}})$, $l_{\text{R2}}$ is the length of Regime 2 \cite{10.1063/5.0278015}. This scaling is performed to establish the generality of Regime 3.

A sharp radial density gradient can be observed at $x/d = 2$ in Fig.~\ref{fig:contour.png}(b) across the jet interface, decreasing $\bar{\rho}_{m}/\bar{\rho}_{s}$ from 2.75 to 1.25 within a radial distance of $\Delta r/d = 0.5$. The radial distance, $\Delta r/d$, of the radial density gradient can be observed to increase along the axial direction; however, the differential ratio $\Delta\bar{\rho}_{m}/\bar{\rho}_{s}\rightarrow 0$. This observation clearly shows a radially expanding but axisymmetric mixing layer along the shear layer and a decaying density of the motive jet along the axial direction. Furthermore, until $x/d\leq 4$ at $-0.25\leq r/d\leq 0.25$ a slight decrease in the mixture density can be observed, and a slight increase when $r/d\rightarrow 1$. These minor changes result from local heat transfer along the wall, which does not significantly affect the flow field in R3.

Developing a reduced-order model using a similarity analysis of the flow field in R3 requires an analytical expression for the radial time-averaged mixture density distribution. We propose to fit the 3D simulation data to a ``sigmoidal function'' of the form,
\begin{equation}
    \label{equ:sample_fit}
    \frac{\bar{\rho}_{m}}{\bar{\rho}_{s}}\left(\frac{X}{d},\frac{r}{d}\right) = a\left[\tanh\left(\frac{2br}{d} + c\right)+\tanh\left(-\frac{2br}{d} + c\right)\right] + e
\end{equation}
In Eq.~\eqref{equ:sample_fit} the coefficients $a,b,c,e$ are all functions of $X/d$. Physically, we can immediately write $a = f[(\bar{\rho}_{m} \ \text{at} \ r/d=0) - (\bar{\rho}_{m} \ \text{at} \ r/d=1)]$ and $e = f(\bar{\rho}_{m} \ \text{at} \ r/d=1)$, as shown in Fig.~\ref{fig:contour.png}(b). The reason that $b,c$ are functions of $X/d$ is the shear layer's expansion in the radial and axial directions. Fitting the 3D simulation data to the analytical form in Eq.~\eqref{equ:sample_fit}, we obtain 
\begin{subequations}
    \label{equ:model_parameters}
    \begin{flalign}
        &a = a(X/d)= -4.918(X/d) + 44.524 && \\
        &b = b(X/d)= -0.5277(X/d) + 4.4276 && \\
        &c = c(X/d)= 0.0681(X/d)^2 - 1.0086(X/d) + 4.0654 && \\
        &e = e(X/d) -0.8187(X/d)^2 + 13.3152(X/d) + 71.6052 &&
    \end{flalign}
\end{subequations}
To confirm the fits, the radial distributions of density at $0 \leq X/d \leq 8$ are plotted in Fig.~\eqref{fig:contour.png}(c) based on Eqs.~\ref{equ:sample_fit} and \eqref{equ:model_parameters}. We observe good agreement between the radial distribution predicted by the costly 3D numerical simulation and the analytical fits in Fig.~\ref{fig:contour.png}(b).

\begin{figure*}[htbp]
    \parbox[c]{\linewidth}{
    \centering
    \begin{subfigure}[htbp]{0.55\linewidth}
        \centering
        \includegraphics[width=1\linewidth]{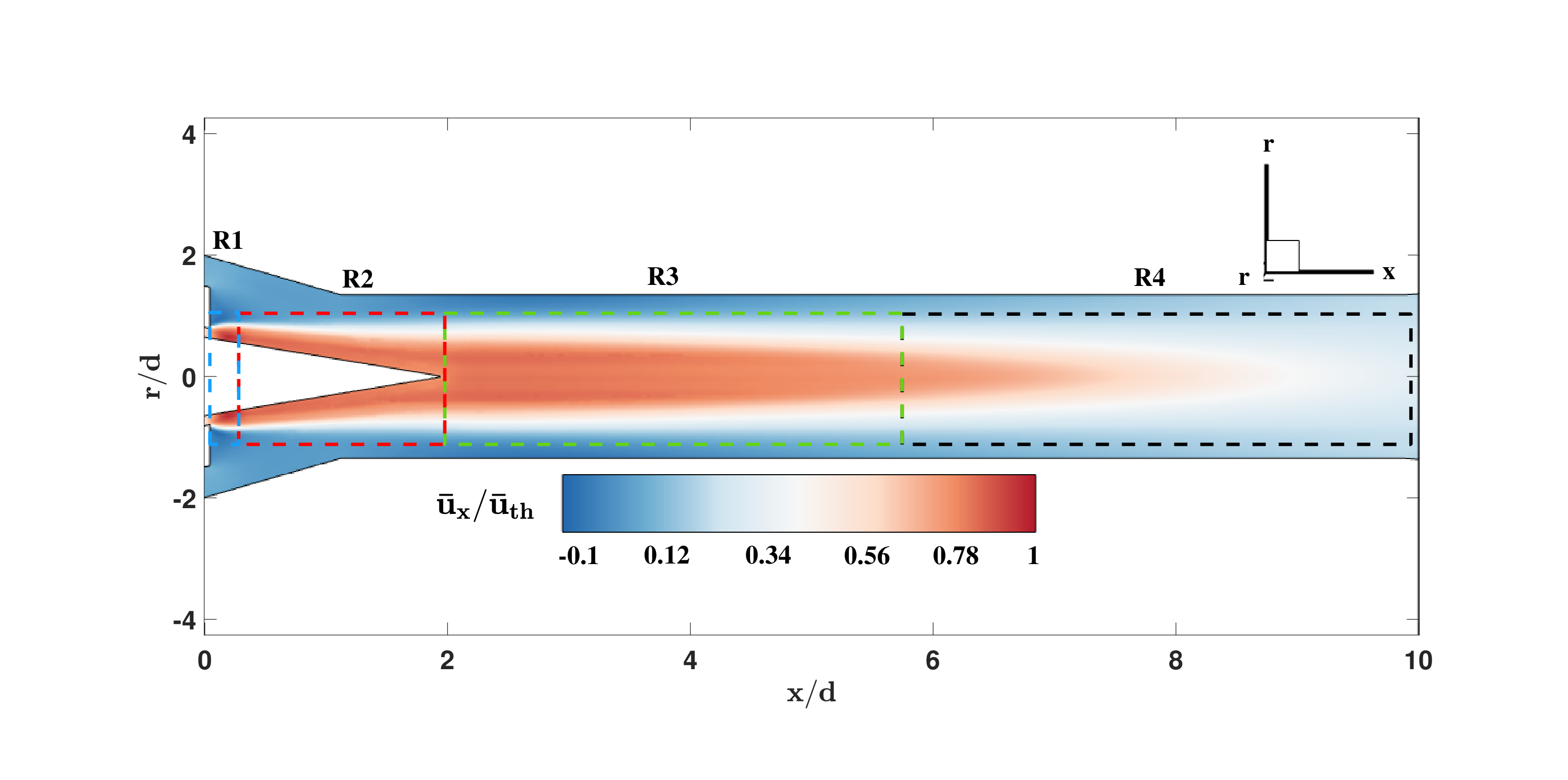}
        \caption{Spatial distribution of $\bar{u}_{x}/\bar{u}_{\text{th}}$}
    \end{subfigure}
    \hfill\hfill
    \begin{subfigure}[htbp]{0.4\linewidth}
        \centering
        \includegraphics[width=1\linewidth]{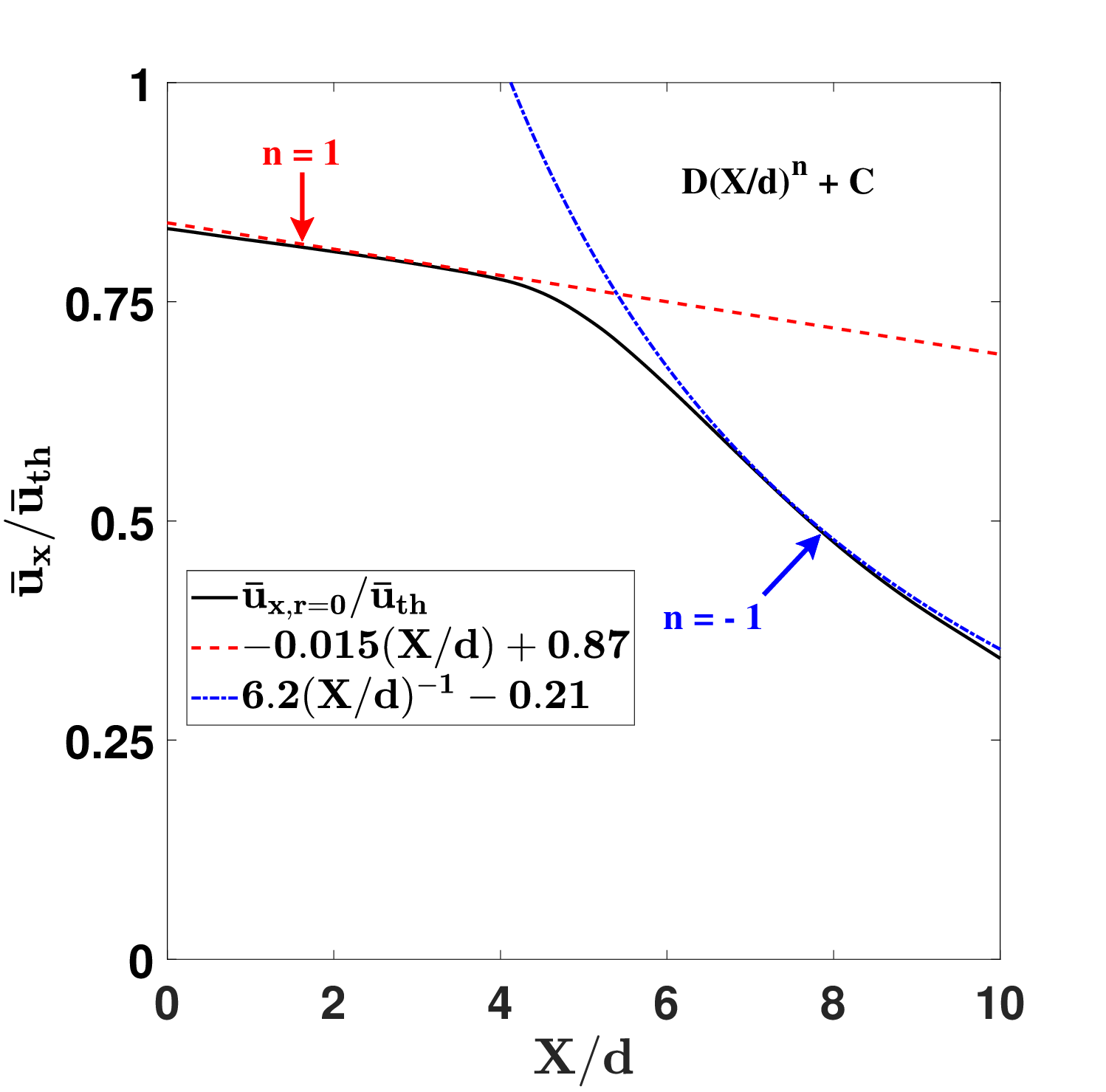}
        \caption{Axial distribution of $\bar{u}_{x}/\bar{u}_{\text{th}}$ at $r/d = 0$}
    \end{subfigure}
    \caption{Spatial and axial distributions of the scaled centerline axial velocity $\bar{u}_{x}/\bar{u}_{\text{th}}$}
    \label{fig:velocity.png}}
\end{figure*}

The primary purpose of ejectors in refrigeration cycles is to increase pressure at the diffuser outlet than that of the suction inlet, namely $P_{d}>P_{s}$. The static pressure at the motive throat, $P_{\text{th}}$, is almost identical to the suction inlet pressure $P_s$ \cite{10.1063/5.0278015}, which means the motive jet experiences a constant adverse pressure gradient ($\partial p/\partial x > 0$) along the axial direction. Figure~\ref{fig:velocity.png}(a) illustrates the spatial distribution of time-averaged axial velocity, $\bar{u}_{x}$, which is represented in a dimensionless form as $\bar{u}_{x}/\bar{u}_{\text{th}}$, where $\bar{u}_{\text{th}}$ is the throat velocity. The velocity at the motive throat of the jet attains a maximum magnitude due to the effect of isothermal expansivity on the bulk velocity through the motive \cite{10.1063/5.0278015}. The centerline velocity, $\bar{u}_{x}$ at $r/d = 0$, of the jet can be seen to decrease axially with the steady radial growth of the shear layer, decaying the potential core of the jet in Fig.~\ref{fig:velocity.png}(a). Quantitative representation of the scaled centerline axial velocity, $\bar{u}_{x}/\bar{u}_{\text{th}}$, at $r/d = 0$, can be seen to have a linear decay following, approximately,
\begin{equation}
    \label{equ:linear_decay}
    \left(\frac{\bar{u}_{x}}{\bar{u}_{\text{th}}}\right)_{r/d = 0} = D_{1}\left(\frac{X}{d}\right) + C_{1} = -0.015\left(\frac{X}{d}\right) + 0.87
\end{equation}
as shown in Fig.~\ref{fig:velocity.png}(b). According to Pope \cite{pope_turbulent_2000}, a linear decay of centerline velocity indicates a decelerating turbulent jet.

A prominent potential core can be observed in R1, R2, and R3, where our previous instantaneous resolved scale model shows that the shear layer is not near-wall \cite{10.1063/5.0278015}. However, the jet in R4 transformed into a wall-bounded turbulent flow, decaying the potential core completely by the wall shear effect as shown in Fig.~\ref{fig:velocity.png}(a). The rate of decay of the potential core due to the wall-shear effect is well described by the fit
\begin{equation}
    \label{equ:nonlinear_decay}
    \left(\frac{\bar{u}_{x}}{\bar{u}_{\text{th}}}\right)_{r/d = 0} = D_{2}\left(\frac{X}{d}\right)^{-1} + C_{2} = 6.2\left(\frac{X}{d}\right)^{-1} - 0.21
\end{equation}
as shown in Fig.~\ref{fig:velocity.png}(b).

The flow acts similarly to a freely expanding jet due to the diverging shape of the diffuser in the downstream flow, following Freund's experiment of a freely expanding jet \cite{freund2001noise}, which is discussed in detail in our previous article \cite{10.1063/5.0278015}.

\section{Similarity Analysis of the Flow Field in Regime 3} 
\label{sec:selfsimilar}

We follow a similarity analysis approach to derive a reduced-order model of the flow field in R3. In this regime, the jet behaves like a negatively buoyant jet, where the effect of gravity is replaced by a constant positive pressure gradient across the ejector, defining an effective inlet Froude number of 14.13 \cite{10.1063/5.0278015}.

The flow field in R3 obeys the conservation of mass and axial momentum transport equations \cite{oud2016fully}, namely
\begin{align}
    \label{equ:mass}
    &\frac{\partial\rho_{m}}{\partial t} + \frac{\partial}{\partial X}(\rho_{m}u_{x}) + \frac{1}{r}\frac{\partial}{\partial r}(r\rho_{m}u_{r}) + \frac{1}{r}\frac{\partial}{\partial\theta}(\rho_{m}u_{\theta}) \nonumber\\ 
    &\qquad = \frac{\partial\rho_{m}}{\partial t} + \vec\nabla\cdot(\rho_{m}\vec{u}) = 0\\
    \label{equ:momentum}
    &\frac{\partial}{\partial t}(\rho_{m}u_{x}) + \frac{\partial}{\partial X}(\rho_{m}u_{x}^{2}) + \frac{1}{r}\frac{\partial}{\partial r}(r\rho_{m}u_{x}u_{r})   \nonumber \\ 
    &+ \frac{1}{r}\frac{\partial}{\partial\theta}(\rho_{m}u_{x}u_{\theta}) = -\frac{\partial p}{\partial X} + \frac{\partial}{\partial X}\left[\mu\left(\frac{\partial u_{x}}{\partial X} - \frac{2}{3}\left(\vec\nabla\cdot\vec{u}\right)\right)\right] \nonumber\\ 
    &+ \frac{1}{r}\frac{\partial}{\partial r}\left[\mu r\left(\frac{\partial u_{x}}{\partial r} + \frac{\partial u_{r}}{\partial X}\right)\right] + \frac{1}{r}\frac{\partial}{\partial\theta}\left[\mu\left(\frac{\partial u_{\theta}}{\partial X} + \frac{1}{r}\frac{\partial u_{x}}{\partial\theta}\right)\right]
\end{align}
The momentum transports in radial and azimuthal directions are neglected.

Previous discussion in section~\ref{sec:flow} shows that the flow is axisymmetric ($\partial/\partial\theta = 0$), isothermal ($\nabla\mu = 0$), and assuming a quasi-steady condition ($\partial/\partial t = 0$) for simplicity, the conservation and transport equations~\eqref{equ:mass} and \eqref{equ:momentum} can be modified as follows:
\begin{align}
    \label{equ:mass_1}
    &\frac{\partial}{\partial X}(\rho_{m}u_{x}) + \frac{\partial}{\partial r}(\rho_{m}u_{r}) + \frac{\rho_{m}u_{r}}{r} = 0\\
    \label{equ:momentum_1}
    &\frac{\partial}{\partial X}(\rho_{m}u_{x}^{2}) + \frac{\partial}{\partial r}(\rho_{m}u_{x}u_{r}) + \frac{\rho_{m}u_{x}u_{r}}{r}  \nonumber \\ 
    &\qquad = -\frac{\partial p}{\partial X} + \mu\left[\frac{1}{3}\frac{\partial^{2}u_{x}}{\partial X^{2}} + \frac{1}{r}\frac{\partial}{\partial r}\left(r\frac{\partial u_{x}}{\partial r}\right)\right]
\end{align}

The jet in the ejector is developed under a high Reynolds number flow, here $\text{Re} \simeq 1.2\times 10^{5}$ \cite{bhaduri2024flow}. This high Reynolds number makes the flow field turbulent with density fluctuations, which requires an appropriate approach to capture the effects of the density fluctuations in the turbulent flow. To this end, we proceed with a time-averaged density-weighted approach, known as Favre averaging \cite{adumitroaie1999improved}. Mathematically, the Favre-averaged quantities (denoted by tildes) are defined as
\begin{equation}
    \label{equ:favre}
    \phi = \tilde{\phi} + \phi'', \qquad \text{where} \quad \tilde{\phi} = \frac{\overline{\rho_{m}\phi}}{\bar{\rho}_{m}}
\end{equation}
We have also given the relation to the Reynolds averaged quantities (denoted by overbars). Observe that $\overline{\rho_{m}\phi''} = 0$ but $\ \overline{\phi''}\neq 0$.

Applying the Favre-averaging \eqref{equ:favre} to the mass and momentum conservation equations \eqref{equ:mass_1} and \eqref{equ:momentum_1}, we obtain
\begin{align}
    \label{equ:mass_2}
    &\frac{\partial}{\partial X}(\bar{\rho}_{m}\tilde{u}_{x}) + \frac{\partial}{\partial r}(\bar{\rho}_{m}\tilde{u}_{r}) + \frac{\bar{\rho}_{m}\tilde{u}_{r}}{r} = 0\\
    \label{equ:momentum_2}
    &\tilde{u}_{x}\frac{\partial\tilde{u}_{x}}{\partial X} + \tilde{u}_{r}\frac{\partial\tilde{u}_{x}}{\partial r} = -\frac{1}{\bar{\rho}_{m}}\frac{\partial\tilde{p}}{\partial X} + \nu\left[\frac{1}{3}\frac{\partial^{2}\tilde{u}_{x}}{\partial X^{2}} + \frac{1}{r}\frac{\partial}{\partial r}\left(r\frac{\partial\tilde{u}_{x}}{\partial r}\right)\right] \nonumber \\ 
    &\qquad- \frac{1}{\bar{\rho}_{m}}\frac{\partial}{\partial X}(\overline{\rho_{m}u_{x}''u_{x}''}) - \frac{1}{r\bar{\rho}_{m}}\frac{\partial}{\partial r}(r\overline{\rho_{m}u_{x}''u_{r}''})
\end{align}

Note that the time-averaged fluctuating  viscous stress term, 
\begin{equation*}
\nu\left[\frac{1}{3}\frac{\partial^{2}\overline{u_{x}''}}{\partial X^{2}} + \frac{1}{r}\frac{\partial}{\partial r}\left(r\frac{\partial\overline{u_{x}''}}{\partial r}\right)\right]
\end{equation*}
was in deriving Eq.~\eqref{equ:momentum_2} because it is negligible for large  Reynolds number flow \cite{adumitroaie1999improved}. In Eq.\ref{equ:momentum_2}, the terms $(1/\bar{\rho}_{m})(\partial/\partial X)(\overline{\rho_{m}u_{x}''u_{x}''})$ and $(1/r\bar{\rho}_{m})(\partial/\partial r)(r\overline{\rho_{m}u_{x}''u_{r}''})$ represents the Reynolds stress terms in axial and radial directions, which are the typical unclosed terms modeled in different turbulence models \cite{pope_turbulent_2000}.

Attempting a similarity transformation of the flow field, $\tilde{u}_{x},\tilde{u}_{r},[\overline{\rho_{m}u_{x}''u_{x}''} + (\tilde{p} - \tilde{p}_{0})]$, and $(\overline{\rho_{m}u_{x}''u_{r}''})$ are represented as a function of the Favre-averaged centerline velocity, $\tilde{u}_{c}$, and shape functions $f(\eta)$, $g(\eta)$, $h(\eta)$, and $k(\eta)$, respectively. Here, $p_{0}$ is a constant pressure, and $\eta$ is a transformed coordinate combining the axial $X$ and radial $r$ coordinates. Based on the discussion in section \ref{sec:flow} (recall Fig.~\ref{fig:velocity.png}(b)), the centerline velocity is represented as
\begin{subequations}
    \label{equ:centerline_velocity}
    \begin{align}
    &\tilde{u}_{c} = D(X/d)^{n} && \\
    &\tilde{u}_{c} =
                    \begin{cases}
                    D(X/d), & \text{ for } 0 \leq X/d \leq 4 \\
                    D(X/d)^{-1}, & \text{ for } X/d \geq 5 
                    \end{cases} &&
    \end{align}
\end{subequations}
The mathematical representations of the transformed variables are then as follows,
\begin{subequations}
    \label{equ:similarity_variables}
    \begin{align}
        &\eta = \frac{r}{X} && \\
        &\tilde{u}_{x} = \tilde{u}_{c}f(\eta) && \\
        &\tilde{u}_{r} = \tilde{u}_{c}g(\eta) && \\
        &\overline{\rho_{m}u_{x}''u_{x}''} + (\tilde{p} - \tilde{p}_{0}) = \tilde{u}_{c}^{2}\bar{\rho}_{m}h(\eta) && \\
        &\overline{\rho_{m}u_{x}''u_{r}''} = \tilde{u}_{c}^{2}\bar{\rho}_{m}k(\eta) &&
    \end{align}
\end{subequations}

Applying the transformations \eqref{equ:similarity_variables} to the mass and momentum conservation equations \eqref{equ:mass_2} and \eqref{equ:momentum_2} and using Eqs.~\eqref{equ:centerline_velocity}, we obtain
\begin{align}
    \label{equ:similarity_mass}
    &g = \frac{1}{\exp\left[\int_{0}^{\eta}\left(\frac{1}{\eta} + \frac{X}{\bar{\rho}_{m}}\frac{\partial\bar{\rho}_{m}}{\partial r}\right)d\eta\right]} \nonumber\\
    &\phantom{g=}\times \int_{0}^{\infty}\exp\left[\int_{0}^{\eta}\left(\frac{1}{\eta} + \frac{X}{\bar{\rho}_{m}}\frac{\partial\bar{\rho}_{m}}{\partial r}\right)d\eta\right] \nonumber\\ 
    &\phantom{g=\int_{0}^{\infty}}\times \left(\eta\frac{df}{d\eta} - \frac{X}{\bar{\rho}_{m}}f\frac{\partial\bar{\rho}_{m}}{\partial X} - nf\right)d\eta\\
    \label{equ:similarity_transformation}
    &nf^{2} - \eta f\frac{df}{d\eta} + g\frac{df}{d\eta} = \eta\frac{dh}{d\eta} - 2nh - \frac{1}{\eta}\frac{d(\eta k)}{d\eta} \nonumber \\ 
    &- \frac{X}{\bar{\rho}_{m}}\left(h\frac{\partial\bar{\rho}_{m}}{\partial X} + k\frac{\partial\bar{\rho}_{m}}{\partial r}\right) \nonumber \\ 
    &+ \frac{1}{\text{Re}_{c}}\left[-\frac{n(1-n)f}{3} + \frac{2}{3}\left(1 - n + \frac{3}{2\eta^{2}}\right)\eta\frac{df}{d\eta} + \frac{(\eta^{2}+3)}{3}\frac{d^{2}f}{d\eta^{2}}\right]
\end{align}
Here, $\text{Re}_{c} = DX^{n+1}/\nu$ represents the transformed centerline Reynolds number, and recall that $f = f(\eta)$, $g = g(\eta)$, $h = h(\eta)$, and $k = k(\eta)$. 

It is clear from Eqs.~\eqref{equ:similarity_mass} and \eqref{equ:similarity_transformation} that the similarity transformation of the conservation equations presents incomplete similarity because $\bar{\rho}_{m}$, which is defined in terms of $\tanh(X,r)$ cannot be expressed as a sole function of $\eta$, remains in the equations. However, at $X/d\geq 7$, $\nabla\bar{\rho}_{m}\rightarrow \vec{0}$ can be transformed into a completely self-similar flow field only when (i) $n = -1$ (to eliminate the dependence of $\text{Re}_c$ on $X$) or (ii) $n = 1$ while simultaneously assuming that $\text{Re}_{c}\rightarrow\infty$.

Either way, we can represent the analytical fit for density, i.e., Eq.~\eqref{equ:sample_fit}, in terms of a new dimensionless similarity variable (namely, $\varepsilon = X/d$) along with transformed coordinate $\eta = r/X$, specifically:
\begin{equation}
    \label{equ:density_analytical}
    \bar{\rho}_{m}(\varepsilon,\eta) = a\left[\tanh\left(2\eta\varepsilon b+c\right) + \tanh\left(-2\eta\varepsilon b+c\right)\right] + e
\end{equation}
where $a,b,c,d$ are now solely functions of $X/d = \varepsilon$. The density gradient in the transformed momentum equation \eqref{equ:similarity_transformation} is then
\begin{multline}
    \label{equ:density_gradient_transformed}
    \frac{X}{\bar{\rho}_{m}}\left(h\frac{\partial\bar{\rho}_{m}}{\partial X} + k\frac{\partial\bar{\rho}_{m}}{\partial r}\right) = \frac{h}{a}\left(1-\frac{e}{\overline{\rho}_{m}}\right)\varepsilon\frac{da}{d\varepsilon}\\ +\frac{2a}{\overline{\rho}_{m}}\varepsilon\left(\eta h\varepsilon\frac{db}{d\varepsilon}+bk\right)\left[\operatorname{sech}^{2}\left(2\eta\varepsilon b+c\right)-\operatorname{sech}^{2}\left(-2\eta\varepsilon b+c\right)\right]\\ +\frac{ah}{\overline{\rho}_{m}}\varepsilon\frac{dc}{d\varepsilon}\left[\operatorname{sech}^{2}\left(2\eta\varepsilon b+c\right)+\operatorname{sech}^{2}\left(-2\eta\varepsilon b+c\right)\right] + \frac{h}{\overline{\rho}_{m}}\varepsilon\frac{de}{d\varepsilon}
\end{multline}
Equation~\eqref{equ:density_gradient_transformed} is a rather complicated expression featuring two variables, which still cannot render  Eqs.~\eqref{equ:similarity_mass} and \eqref{equ:similarity_transformation} completely self-similar. However, having introduced $\varepsilon$, we can now consider an asymptotic expansion approach with $\varepsilon = X/d\ll1$ as the perturbation parameter, yielding a sequence of locally self-similar problems out of Eqs.~\eqref{equ:similarity_mass} and \eqref{equ:similarity_transformation}.

\begin{figure}[htbp]
    \centering
    \includegraphics[width=\linewidth]{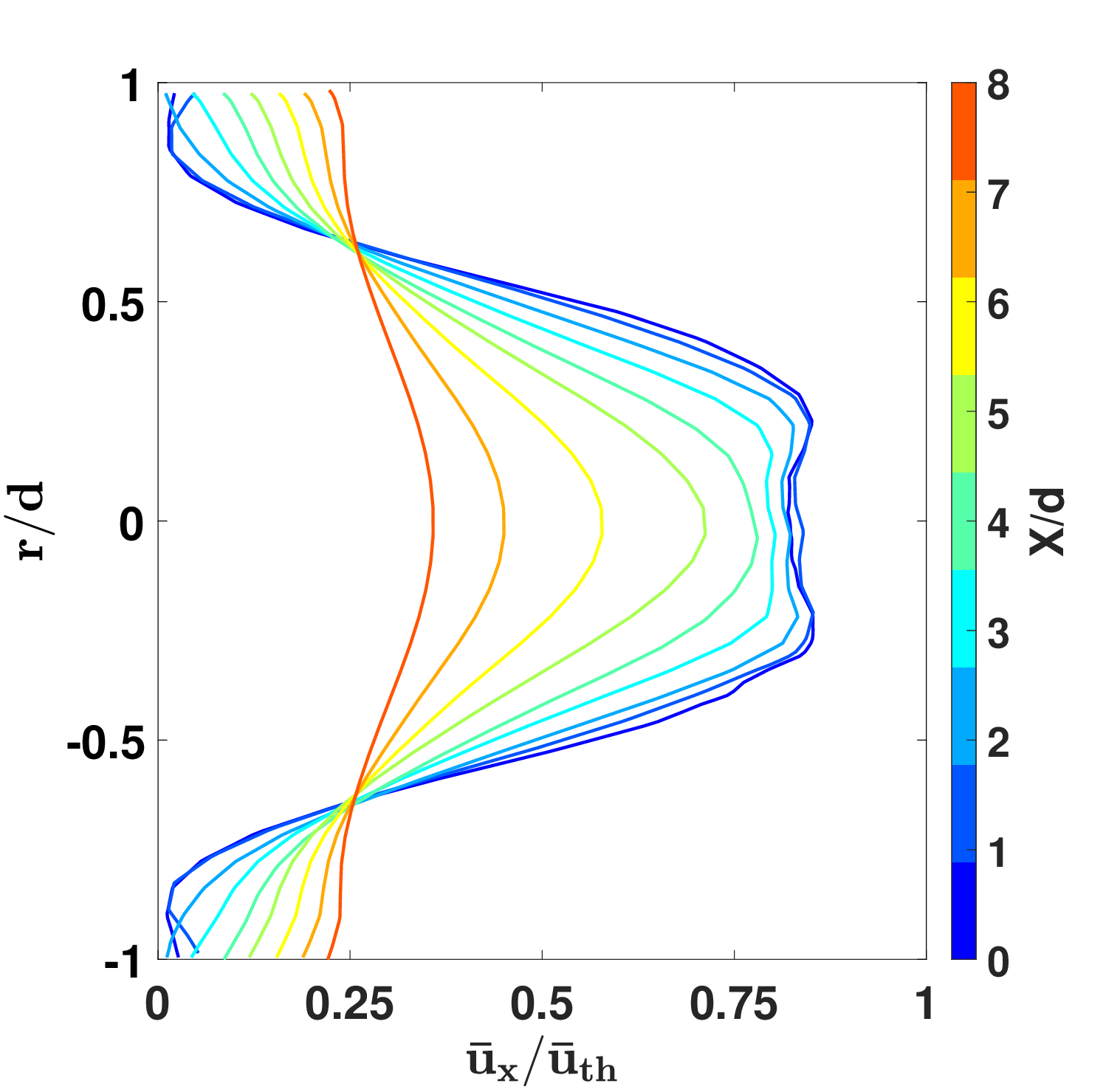}
    \caption{\label{fig:velocity_rad.png}Radial distribution of $\bar{u}_{x}/\bar{u}_{\text{th}}$ at $2\leq x/d\leq 10$}
\end{figure}

The scaled axial velocity distribution, $\bar{u}_{x}/\bar{u}_{\text{th}}$ at $2\leq x/d\leq 10$, is shown in Fig.~\ref{fig:velocity_rad.png}. The radial gradient of $\bar{u}_{x}/\bar{u}_{\text{th}}$ can be seen to be decreasing downstream, due to the acceleration of the quiescent gaseous fluid surrounding the `motive' jet. Furthermore, the velocity of the jet at $|r/d|\leq 0.5$ decreases downstream by momentum diffusion triggered by mixing along the shear layer. Mixing decreases the inertial difference between the surrounding gas and the jet, leading towards a radial equilibrium of kinetic energy. However, the prior effect of the boundary layer by the converging needle can be observed in the velocity distribution at $x/d\leq 5$, showing a slight decrease near $r/d = 0$.

Some mathematical manipulation of Eq.~\eqref{equ:similarity_mass} provides more information on the approximate nature of the solution. Specifically, we can write
\begin{equation}
    \label{equ:nature_mass}
    g = \frac{1}{G}\int_{0}^{\infty}G\left(\eta\frac{df}{d\eta}-\frac{\varepsilon d}{\overline{\rho}_{m}}fF\left[\operatorname{tanh}(\varepsilon,\eta)\right]-nf\right)d\eta
\end{equation}
where
\begin{equation*}
    G = \left[\eta+\int\exp\left(\frac{\varepsilon d}{\overline{\rho}_{m}}F\left[\operatorname{tanh}(\varepsilon,\eta)\right]\right)d\eta\right]
\end{equation*}
and $\lim_{\eta\rightarrow\infty}(G/\eta) = 1 + \exp(\varepsilon d/\overline{\rho}_{m})$, which is a constant. The second integral is a function of the density gradient, which leads to incompleteness in the self-similarity reduction and requires further attention. Its nature is similar to $\tanh(\varepsilon,\eta)$ using asymptotic approximation. Furthermore, the presence of $\tanh(\varepsilon,\eta)$ in the expression for $g(\eta)$ also influences the transformed momentum equation Eq.~\eqref{equ:similarity_transformation}. Indeed, the effect of the density gradient is confirmed by observing the radial velocity distribution in Fig.~\ref{fig:velocity_rad.png}.

\section{Conclusions} 
\label{sec:conclusions}

This work discusses the development of a reduced-order model of R3 inside ejectors. The flow field inside the R3 shows an incomplete (or, perhaps, local) self-similarity behavior due to the sharp density gradient of the flow. A self-similar reduction can be achieved when $\vec\nabla\bar{\rho}_{m}\rightarrow \vec{0}$, which occurs for $X/d\geq 7$, near the end of R3. Therefore, towards the beginning of R3, the self-similar behavior may be studied using asymptotic expansion with $\varepsilon = X/d\ll1$ as the perturbation parameter. The following conclusions can be drawn based on the preceding analysis:
\begin{itemize}
    \item The flow field in R3 inside the motive jet of an ejector possesses a distinct shear layer with a sharp radial density gradient.
    \item The sharp radial density gradient gradually homogenizes downstream by mixing across the shear layer.
    \item We developed an analytical fit capturing the axial and radial mixture density gradient variations.
    \item We also demonstrated that another analytical fit captures the axial variations of centerline velocity, which is a linearly decreasing function ($\sim X$) until R3, followed by a hyperbolic decay ($\sim X^{-1}$) afterwards, behaving as a decelerating jet before transforming into a wall-bounded expanding jet.
    \item A similarity transformation of the Favre-averaged mass and momentum conservation equations of the flow field in R3 shows an incomplete self-similarity due to the sharp density gradient across the shear layer (a distinct turbulent interface) in a sufficiently high Reynolds number flow when $n = 1$ or for any Reynolds numbers when $n = -1$.
    \item We hypothesize that the solution of the transformed momentum transport equation can be found in terms of the transformed coordinate $\eta = r/X$ by utilizing an asymptotic expansion in terms of the perturbation parameter $\varepsilon = X/d$.
\end{itemize}

\section*{Appendix}
\noindent The conservation of mass and axial momentum transport equations are \cite{oud2016fully},
\begin{align}
    \label{equ:mass}
    &\frac{\partial\rho_{m}}{\partial t} + \frac{\partial}{\partial X}(\rho_{m}u_{x}) + \frac{1}{r}\frac{\partial}{\partial r}(r\rho_{m}u_{r}) + \frac{1}{r}\frac{\partial}{\partial\theta}(\rho_{m}u_{\theta}) \nonumber\\ 
    &\qquad = \frac{\partial\rho_{m}}{\partial t} + \vec\nabla\cdot(\rho_{m}\vec{u}) = 0\\
    \label{equ:momentum}
    &\frac{\partial}{\partial t}(\rho_{m}u_{x}) + \frac{\partial}{\partial X}(\rho_{m}u_{x}^{2}) + \frac{1}{r}\frac{\partial}{\partial r}(r\rho_{m}u_{x}u_{r})   \nonumber \\ 
    &+ \frac{1}{r}\frac{\partial}{\partial\theta}(\rho_{m}u_{x}u_{\theta}) = -\frac{\partial p}{\partial X} + \frac{\partial}{\partial X}\left[\mu\left(\frac{\partial u_{x}}{\partial X} - \frac{2}{3}\left(\vec\nabla\cdot\vec{u}\right)\right)\right] \nonumber\\ 
    &+ \frac{1}{r}\frac{\partial}{\partial r}\left[\mu r\left(\frac{\partial u_{x}}{\partial r} + \frac{\partial u_{r}}{\partial X}\right)\right] + \frac{1}{r}\frac{\partial}{\partial\theta}\left[\mu\left(\frac{\partial u_{\theta}}{\partial X} + \frac{1}{r}\frac{\partial u_{x}}{\partial\theta}\right)\right]
\end{align}
The momentum transports in radial and azimuthal directions are neglected. In case of an axisymmetric ($\partial/\partial\theta = 0$), isothermal ($\nabla\mu = 0$), and a quasi-steady condition ($\partial/\partial t = 0$) flow conditions, the conservation and transport equations~\eqref{equ:mass} and \eqref{equ:momentum} are modified as follows:
\begin{align}
    \label{equ:mass_1}
    &\frac{\partial}{\partial X}(\rho_{m}u_{x}) + \frac{\partial}{\partial r}(\rho_{m}u_{r}) + \frac{\rho_{m}u_{r}}{r} = 0\\
    \label{equ:momentum_1}
    &\frac{\partial}{\partial X}(\rho_{m}u_{x}^{2}) + \frac{\partial}{\partial r}(\rho_{m}u_{x}u_{r}) + \frac{\rho_{m}u_{x}u_{r}}{r}  \nonumber \\ 
    &\qquad = -\frac{\partial p}{\partial X} + \mu\left[\frac{1}{3}\frac{\partial^{2}u_{x}}{\partial X^{2}} + \frac{1}{r}\frac{\partial}{\partial r}\left(r\frac{\partial u_{x}}{\partial r}\right)\right]
\end{align}
The Favre-averaged quantities (denoted by tildes) are defined as,
\begin{subequations}
    \label{equ:favre}
    \begin{align}
    &\phi = \tilde{\phi} + \phi'', \qquad \text{where} \quad \tilde{\phi} = \frac{\overline{\rho_{m}\phi}}{\bar{\rho}_{m}} && \\
    &\overline{\rho_{m}\phi''} = 0 && \\
    &\overline{\phi''}\neq 0 &&
    \end{align}
\end{subequations}
The Reynolds averaged quantities are denoted by overbars. The Reynolds averaged conservation of mass from equation~\eqref{equ:mass_1} is,
\begin{multline}
    \label{equ:mass_ra_1}
    \overline{\frac{\partial}{\partial X}(\rho_{m}\tilde{u}_{x}+\rho_{m}u_{x}^{''})} + \overline{\frac{\partial}{\partial r}(\rho_{m}\tilde{u}_{r}+\rho_{m}u_{r}^{''})} + \overline{\frac{\rho_{m}\tilde{u}_{r}+\rho_{m}u_{r}^{''}}{r}} = 0 \\ \implies \frac{\partial}{\partial X}(\overline{\rho_{m}\tilde{u}_{x}}) + \frac{\partial}{\partial r}(\overline{\rho_{m}\tilde{u}_{r}}) + \frac{\overline{\rho_{m}\tilde{u}_{r}}}{r} + \frac{\partial}{\partial X}(\overline{\rho_{m}u_{x}^{''}}) + \frac{\partial}{\partial r}(\overline{\rho_{m}u_{r}^{''}}) \\ + \frac{\overline{\rho_{m}u_{r}^{''}}}{r} = 0 \\ \therefore \frac{\partial}{\partial X}(\overline{\rho}_{m}\tilde{u}_{x}) + \frac{\partial}{\partial r}(\overline{\rho}_{m}\tilde{u}_{r}) + \frac{\overline{\rho}_{m}\tilde{u}_{r}}{r} = 0
\end{multline}
The Reynolds averaged momentum transport equation from~\eqref{equ:mass_1} is,
\begin{multline}
    \label{equ:moment_ra_1}
    \overline{\frac{\partial}{\partial X}[\rho_{m}(\tilde{u}_{x}+u_{x}^{''})^{2}]} + \overline{\frac{\partial}{\partial r}[\rho_{m}(\tilde{u}_{x}+u_{x}^{''})(\tilde{u}_{r}+u_{r}^{''})]} \\ + \overline{\frac{\rho_{m}(\tilde{u}_{x}+u_{x}^{''})(\tilde{u}_{r}+u_{r}^{''})}{r}} = -\overline{\frac{\partial (\tilde{p}+p^{''})}{\partial X}}\\ + \overline{\mu\left[\frac{1}{3}\frac{\partial^{2}(\tilde{u}_{x}+u_{x}^{''})}{\partial X^{2}} + \frac{1}{r}\frac{\partial}{\partial r}\left(r\frac{\partial (\tilde{u}_{x}+u_{x}^{''})}{\partial r}\right)\right]} \\ \implies \tilde{u}_{x}\frac{\partial\tilde{u}_{x}}{\partial X} + \frac{1}{\bar{\rho}_{m}}\frac{\partial}{\partial X}(\overline{\rho_{m}u_{x}''u_{x}''}) + \tilde{u}_{r}\frac{\partial\tilde{u}_{x}}{\partial r} + \frac{1}{r\bar{\rho}_{m}}\frac{\partial}{\partial r}(r\overline{\rho_{m}u_{x}''u_{r}''})\\ = -\frac{1}{\bar{\rho}_{m}}\frac{\partial\tilde{p}}{\partial X} + \frac{\mu}{\overline{\rho}_{m}}\left[\frac{1}{3}\frac{\partial^{2}\tilde{u}_{x}}{\partial X^{2}} + \frac{1}{r}\frac{\partial}{\partial r}\left(r\frac{\partial\tilde{u}_{x}}{\partial r}\right)\right] \\ \therefore \tilde{u}_{x}\frac{\partial\tilde{u}_{x}}{\partial X} + \tilde{u}_{r}\frac{\partial\tilde{u}_{x}}{\partial r} = -\frac{1}{\bar{\rho}_{m}}\frac{\partial\tilde{p}}{\partial X} + \nu\left[\frac{1}{3}\frac{\partial^{2}\tilde{u}_{x}}{\partial X^{2}} + \frac{1}{r}\frac{\partial}{\partial r}\left(r\frac{\partial\tilde{u}_{x}}{\partial r}\right)\right] \\ - \frac{1}{\bar{\rho}_{m}}\frac{\partial}{\partial X}(\overline{\rho_{m}u_{x}''u_{x}''}) - \frac{1}{r\bar{\rho}_{m}}\frac{\partial}{\partial r}(r\overline{\rho_{m}u_{x}''u_{r}''})
\end{multline}
Hence, the Favre-averaged axisymmetric, quasi-steady mass and momentum conservation equations are,
\begin{align}
    \label{equ:mass_2}
    &\frac{\partial}{\partial X}(\bar{\rho}_{m}\tilde{u}_{x}) + \frac{\partial}{\partial r}(\bar{\rho}_{m}\tilde{u}_{r}) + \frac{\bar{\rho}_{m}\tilde{u}_{r}}{r} = 0\\
    \label{equ:momentum_2}
    &\tilde{u}_{x}\frac{\partial\tilde{u}_{x}}{\partial X} + \tilde{u}_{r}\frac{\partial\tilde{u}_{x}}{\partial r} = -\frac{1}{\bar{\rho}_{m}}\frac{\partial\tilde{p}}{\partial X} + \nu\left[\frac{1}{3}\frac{\partial^{2}\tilde{u}_{x}}{\partial X^{2}} + \frac{1}{r}\frac{\partial}{\partial r}\left(r\frac{\partial\tilde{u}_{x}}{\partial r}\right)\right] \nonumber \\ 
    &\qquad- \frac{1}{\bar{\rho}_{m}}\frac{\partial}{\partial X}(\overline{\rho_{m}u_{x}''u_{x}''}) - \frac{1}{r\bar{\rho}_{m}}\frac{\partial}{\partial r}(r\overline{\rho_{m}u_{x}''u_{r}''})
\end{align}
Here, $X = (x-x_{0})$ is the shifted axial coordinate. The mathematical representations of the transformed variables are then as follows,
\begin{subequations}
    \label{equ:similarity_variables}
    \begin{align}
        &\eta = \frac{r}{X} && \\
        &\tilde{u}_{x} = \tilde{u}_{c}f(\eta) && \\
        &\tilde{u}_{r} = \tilde{u}_{c}g(\eta) && \\
        &\overline{\rho_{m}u_{x}''u_{x}''} + (\tilde{p} - \tilde{p}_{0}) = \tilde{u}_{c}^{2}\bar{\rho}_{m}h(\eta) && \\
        &\overline{\rho_{m}u_{x}''u_{r}''} = \tilde{u}_{c}^{2}\bar{\rho}_{m}k(\eta) &&
    \end{align}
\end{subequations}
Here $\tilde{u}_{c}$ is the centerline Favre-averaged velocity, which according to Pope \cite{pope_turbulent_2000}, is
\begin{subequations}
    \label{equ:centerline_velocity}
    \begin{align}
    &\tilde{u}_{c} = D(X/d)^{n} && \\
    &\tilde{u}_{c} =
                    \begin{cases}
                    D(X/d), & \text{ for } 0 \leq X/d \leq 4 \\
                    D(X/d)^{-1}, & \text{ for } X/d \geq 5 
                    \end{cases} &&
    \end{align}
\end{subequations}
Some of the key mathematically transformed expressions are as follows,
\begin{subequations}
    \label{equ:key_expressions}
    \begin{align}
        &\tilde{u}_{x}\frac{\partial\overline{\rho}_{m}}{\partial X} = \tilde{u}_{c}f\frac{\partial\overline{\rho}_{m}}{\partial X} && \\
        &\overline{\rho}_{m}\frac{\partial\tilde{u}_{x}}{\partial X} = \frac{\overline{\rho}_{m}\tilde{u}_{c}}{X}\left(nf - \eta\frac{df}{d\eta}\right) && \\
        &\tilde{u}_{r}\frac{\partial\overline{\rho}_{m}}{\partial r} = \tilde{u}_{c}g\frac{\partial\overline{\rho}_{m}}{\partial r} && \\
        &\overline{\rho}_{m}\frac{\partial\tilde{u}_{r}}{\partial r} = \frac{\overline{\rho}_{m}\tilde{u}_{c}}{X}\frac{dg}{d\eta} && \\
        &\frac{\overline{\rho}_{m}\tilde{u}_{r}}{r} = \frac{\overline{\rho}_{m}\tilde{u}_{c}}{X}\frac{g}{\eta} && \\
        &\tilde{u}_{x}\frac{\partial\tilde{u}_{x}}{\partial X} = \frac{\tilde{u}_{x}^{2}}{X}\left(nf^{2}-\eta f\frac{df}{d\eta}\right) && \\
        &\tilde{u}_{r}\frac{\partial\tilde{u}_{x}}{\partial r} = \frac{\tilde{u}_{c}^{2}}{X}g\frac{df}{d\eta} && \\
        &\frac{\partial}{\partial X}\left[\frac{\overline{\rho_{m}u_{x}''u_{r}''}}{\overline{\rho}_{m}} + \frac{(\tilde{p} - \tilde{p}_{0})}{\overline{\rho}_{m}}\right] = \frac{\tilde{u}_{x}^{2}}{X}\left(2nh + \frac{X}{\overline{\rho_{m}}}h\frac{\partial\overline{\rho}_{m}}{\partial X} - \eta\frac{dh}{d\eta}\right) && \\
        &\frac{1}{r\overline{\rho}_{m}}\frac{\partial}{\partial r}\left(r\overline{\rho_{m}u_{x}''u_{r}''}\right) = \frac{\tilde{u}_{c}^{2}}{X}\left[\frac{1}{\eta}\frac{d}{d\eta}(\eta h) + \frac{X}{\overline{\rho}_{m}}k\frac{\partial\overline{\rho}_{m}}{\partial r}\right] && \\
        &\frac{1}{3}\frac{\partial^{2}\tilde{u}_{x}}{\partial X^{2}} = \frac{\tilde{u}_{c}}{X^{2}}\left[\frac{n(n-1)}{3}f - \frac{2}{3}(n-1)\eta\frac{df}{d\eta} + \frac{\eta^{2}}{3}\frac{d^{2}f}{d\eta^{2}}\right] && \\
        &\frac{1}{r}\frac{\partial}{\partial r}\left(r\frac{\partial\tilde{u}_{x}}{\partial r}\right) = \frac{\tilde{u}_{c}}{X^{2}}\left(\frac{1}{\eta}\frac{df}{d\eta} + \frac{d^{2}f}{d\eta^{2}}\right) &&
    \end{align}
\end{subequations}
Transformation of equation~\eqref{equ:mass_2} using the transformed expressions~\eqref{equ:key_expressions},
\begin{equation}
    \label{equ:mass_t1}
    \begin{aligned}
    \frac{dg}{d\eta} + \left(\frac{1}{\eta} + \frac{X}{\overline{\rho}_{m}}\frac{\partial\overline{\rho}_{m}}{\partial r}\right)g &= \eta\frac{df}{d\eta} - \frac{X}{\overline{\rho}_{m}}f\frac{\partial\overline{\rho}_{m}}{\partial X} - nf \\ &= g' + P(\eta)g \\ &= Q(\eta),
    \end{aligned}
\end{equation}
when $\varepsilon={X}/{d}=\text{constant}$ due to local similarity.
The transformed equation~\eqref{equ:mass_t1} can be solved using an integration factor (IF), which is defined as follows,
\begin{equation}
    \label{equ:if}
    \begin{aligned}
    \text{IF} &= \exp\int P(\eta)d\eta \\
    &= \exp\int\left(\frac{1}{\eta} + \frac{X}{\overline{\rho}_{m}}\frac{\partial\overline{\rho}_{m}}{\partial r}\right)d\eta\\ 
    &= \exp(\log_{e}\eta) + \exp\int\frac{X}{\overline{\rho}_{m}}\frac{\partial\overline{\rho}_{m}}{\partial r}d\eta \\
    &= \eta + \exp\int\frac{X}{\overline{\rho}_{m}}F\left[\operatorname{tanh}(X,\eta)\right]d\eta
    \end{aligned}
\end{equation}
Therefore, the shape function $g(\eta)$ for the radial velocity component is,
\begin{multline}
    \label{equ:nature_mass}
    g = \frac{1}{G}\int_{0}^{\infty}G\left(\eta\frac{df}{d\eta}-\frac{X}{\overline{\rho}_{m}}fF\left[\operatorname{tanh}(X,\eta)\right]-nf\right)d\eta
\end{multline}
where
\begin{equation*}
    G = \left[\eta+\int\exp\left(\frac{X}{\overline{\rho}_{m}}F\left[\operatorname{tanh}(X,\eta)\right]\right)d\eta\right]
\end{equation*}
Transformation of momentum equation in~\eqref{equ:momentum_2} is as follows,
\begin{multline}
    \label{equ:momentum_t1}
    nf^{2} - \eta f\frac{df}{d\eta} + g\frac{df}{d\eta} = \eta\frac{dh}{d\eta} - 2nh - \frac{1}{\eta}\frac{d(\eta k)}{d\eta} \nonumber \\ 
    - \frac{X}{\bar{\rho}_{m}}\left(h\frac{\partial\bar{\rho}_{m}}{\partial X} + k\frac{\partial\bar{\rho}_{m}}{\partial r}\right) \nonumber \\ 
    + \frac{\nu}{\tilde{u}_{c}x}\left[-\frac{n(1-n)f}{3} + \frac{2}{3}\left(1 - n + \frac{3}{2\eta^{2}}\right)\eta\frac{df}{d\eta} + \frac{(\eta^{2}+3)}{3}\frac{d^{2}f}{d\eta^{2}}\right]
\end{multline}
Here, $\text{Re}_{c} = \tilde{u}_{c}x/\nu = DX^{n+1}/\nu$ represents the transformed centerline Reynolds number.

\section*{Declaration of Competing Interest}
\noindent The authors declare that they have no known competing financial interests or personal relationships that could have appeared to influence the work reported in this paper.

\section*{Funding Acknowledgments}
\noindent This research has been funded by Bechtel Corporation, US.

\section*{Author's details}
\noindent \textbf{Sreetam Bhaduri}  \\
Graduate Research Assistant \\
School of Mechanical Engineering\\
Purdue University, West Lafayette, Indiana – 47907, USA\\
Email: \href{mailto:bhaduri@purdue.edu}{bhaduri@purdue.edu}\\
ORCiD: \url{https://orcid.org/0000-0002-5201-3976}\\
\\
\textbf{Ivan C. Christov} \\
Associate Professor\\
School of Mechanical Engineering\\
Purdue University, West Lafayette, Indiana – 47907, USA\\
Email: \href{mailto:christov@purdue.edu}{christov@purdue.edu}\\
ORCiD: \url{https://orcid.org/0000-0001-8531-0531}\\
\\
\textbf{Eckhard A. Groll} \\
William E. and Florence E. Perry Head of Mechanical Engineering, and Reilly Distinguished Professor of Mechanical Engineering\\
Purdue University, West Lafayette, Indiana – 47907, USA\\
Email: \href{mailto:groll@purdue.edu}{groll@purdue.edu}\\
ORCiD: \url{https://orcid.org/0000-0002-1034-7089}\\
\\
\textbf{Davide Ziviani} \\
Associate Professor\\
School of Mechanical Engineering\\
Purdue University, West Lafayette, Indiana – 47907, USA\\
Email: \href{mailto:dziviani@purdue.edu}{dziviani@purdue.edu}\\
ORCiD: \url{https://orcid.org/0000-0002-1778-445X}\\

\bibliographystyle{asmeconf}
\bibliography{asmeconf-sample}
\end{document}